\documentclass{iopjournal}

\usepackage{amsmath, amssymb}
\usepackage{subfig}
\usepackage{subcaption}

\makeatletter
\renewcommand{\p@subfigure}{\thefigure}
\makeatother
\usepackage{multirow}
\usepackage{array}
\usepackage{xspace}
\usepackage{bm}
\usepackage{url}
\usepackage[table]{xcolor}
\usepackage[normalem]{ulem}
\usepackage[inline]{asymptote}
\usepackage{amsmath}
\usepackage{caption}
\usepackage[export]{adjustbox}
\usepackage{subfig}
\usepackage{tikz}
\usepackage{graphicx}
\usepackage[dvipsnames]{xcolor}
\graphicspath{{Figures}}

\begin{document}


\title{Radar Cross-Section Reduction of the Nozzle of an Airborne Platform Using Lightweight Auxetic Metamaterials}

\author{A. Phanendra Kumar$^1$\orcid{0000-0002-7717-0599}, Preeti Kumari$^2$, Dineshkumar Harursampath$^1$\orcid{0000-0001-6855-303X} and Vijay Kumar Sutrakar$^{2,*}$\orcid{0000-0001-9186-5816}}

\affil{$^1$NMCAD Laboratory, Department of Aerospace Engineering, Indian Institute of Science, Bengaluru, Karnataka, India}

\affil{$^2$Aeronautical Development Establishment, DRDO, Bengaluru, India}

\affil{$^*$Author to whom any correspondence should be addressed.}

\email{vks.ade@gov.in}


\begin{abstract}
The nozzle of an aircraft is a major source of radar scattering from the rear aspect of the aircraft, which undergoes higher operational temperatures. In order to reduce the radar scattering of these nozzles, high temperature radar absorbing materials (RAM) are essential. The thickness of these RAM typically increases to attain RCS reduction at lower frequencies, which subsequently leads to a higher weight of the structure. Therefore, this research study investigates the weight advantages of a star auxetic (SA) lattice made up of barium titanate (BaTiO$_{3}$) to reduce the RCS of aircraft exhaust nozzles in the frequency range of 8-18 GHz.  Modelling of SA with a complicated aircraft structure may lead to complexities in terms of Computer Aided Design (CAD)/electromagnetic modelling and higher computational time for solving the electromagnetic problem using exact solvers. In order to simplify the computational problem, a homogenisation and modified transfer matrix method is used to generate the RL performance. 
The RL from the proposed in-house tools is also compared with the Floquet port analysis. 
The RL performance obtained from the proposed method is also validated against experimental data.
Comparative analyses are performed between SA and solid pure-block (PB) BaTiO$_{3}$ samples over 32,761 (181 $\times$ 181) SA–PB thickness combinations. Results show that selected SA samples with the same thickness achieve a minimum RL up to -20 dB lower and a bandwidth of 2–4 GHz broader than PB, leading to a weight saving of approximately 60\%. The median RCS of the nozzle rear aspect also indicates that the SA-based BaTiO$_3$ has an advantage in terms of weight penalty with similar or better RCS performance.
The study demonstrates that auxetic metamaterials will be a multifunctional, lightweight, thermally stable, and radar-absorbent structure for high-temperature aircraft applications.
\end{abstract}

\section{Introduction}
\label{sec: introduction}
Among the significant contributors to the aircraft's radar signature are the cavities, such as the exhaust nozzle in the rear aspect \cite{kim2025development}. 
The internal nozzle areas from the tail-on aspect act as a significant hotspot region, as they tend to scatter the incident radar waves, which often significantly contribute to an aircraft's total tail-on RCS.
The exhaust nozzle will be subjected to high thermal temperatures due to the exhaust gases from the engine.
Harsh thermal conditions at the nozzle areas require materials that can operate under long-term exposure to high temperatures without any degradation in the RL performance. 
Therefore, developing effective radar-absorbing materials for these specific cavities is a critical engineering challenge.
From a historical perspective, the evolution of radar technology has been an epitome of modern aerospace and defence system development. The advancements in radar technology are growing day by day \cite{li2023joint, zhang2025oscjc, bajuri2025recent}. 
This leads to an increasing need for materials that can absorb radar waves efficiently and thus reduce the RCS of military systems, along with the external electromagnetic shaping.
As a result, research on high-performance microwave absorbers has evolved into a strategic necessity in the effort to ensure the low observability and survivability of future aircraft, missiles, and unmanned aerial vehicles \cite{ray2024advances}.
To meet this strategic need for durable, high-performance absorbers, particularly in extreme environments, researchers have increasingly focused on different material classes.
Ceramic materials have been one of the probable candidates for researchers over the past decade due to their unique electromagnetic and thermal properties at higher temperatures \cite{li2022principles, li2023ceramic}.
Many perovskite-structured ceramics exhibit strong interactions with electromagnetic radiation at microwave frequencies, such as high permittivity and dielectric loss mechanisms \cite{khirade2022perovskite}.
Barium titanate (BaTiO$_{3}$) is one of the perovskite structures which can highly absorb the incident radar energy through the dielectric loss mechanism due to its high electrical permittivity and intrinsic polarisation capabilities \cite{gao2011theoretical, wen2015effect, huang2021effect}. 
In addition to their radar-absorbing capabilities, ceramic materials exhibit significantly enhanced thermal stability, a crucial property required for nozzle applications \cite{zhao2023overview, wyatt2024ultra}.

Even though ceramics are a probable candidate material for radar absorption applications, there are still challenges associated with using monolithic ceramic blocks as radar absorbers \cite{eswara2016monolithic}.
The main drawbacks of ceramics are their inherent brittle nature and high density \cite{flower2012high}.
These characteristics of the ceramics have limited the possibility of their practical deployment in aerospace applications, where the weight and mechanical flexibility of the structure are critical.
Auxetic materials are a class of mechanical metamaterials that exhibit a negative Poisson's ratio.
Materials with a negative Poisson ratio expand longitudinally when the structure is expanded laterally, providing the mechanical flexibility \cite{li2024experimental}.
These auxetics are explored limitedly in the literature for their microwave absorption capabilities \cite{damian2007emc, lim2014electromagnetic, tao2022novel}.
In order to overcome the weight penalty and electromagnetic performance, the proposed work investigates the potential of lightweight auxetic metamaterials.
The selective use of auxetic architectures is a compelling technique to overcome the drawbacks, like weight and brittleness, of homogeneous ceramic monolithic structures \cite{kumar2023computational, kumar2024novel}. 
Different ceramic-based auxetic configurations have been studied in the literature, and it is reported that the star-based geometry exhibited superior EM absorption owing to its enhanced incident EM field interaction and impedance matching. Compared to other auxetic configurations, the star configuration has achieved higher RL and broader absorption bandwidth, making it more effective for stealth applications \cite{kumar2023computational}. Therefore, the SA geometric configuration used in the research work carried out by Kumar et al. \cite{kumar2023computational} has been considered in the present study.

Unlike monolithic ceramics, auxetic metamaterial-based lattices can be engineered to balance mechanical robustness \cite{evans2000auxetic, etemadi2024enhancing, li2024compressive} and electromagnetic absorption performance.
With auxetic metamaterials, multiple scattering centres can be introduced along with additional resonance pathways due to the open cell structure \cite{yan2024tunable}.
Due to these additional attributes these open cell structures provide, the microwave absorption bandwidth can be increased along with minimising RL \cite{kumar2023computational}.
This open-cell structure can be tailored to absorb the radar waves incident from multiple directions, ensuring angular independent microwave absorption characteristics \cite{shi2020dispersion}. 
Additionally, even though dense ceramic materials like BaTiO$_{3}$ (BTO) are used for designing the auxetic metamaterials, the open-cell structure aids in reducing the overall weight of the structure compared with the monolithic structures. 
In the current work, the RCS of the exhaust nozzle with a ceramic-based auxetic layer made up of BTO nanoparticles has been studied in the 8 to 18 GHz frequency range, i.e., the X and Ku bands. Other ceramic-based radar absorbent metamaterials, such as SiC honeycomb \cite{wang2024material}, Al$_2$O$_3$/SiC honeycomb \cite{mei20193d}, SiCN honeycomb \cite{pan2022high},  Al$_2$O$_3$/SiC$_{\text{nw}}$/SiOC composite-based metamaterial \cite{mei20213d}, have also been reported in the literature in recent times for high-temperature applications. In general, lattice geometries can be designed to seamlessly integrate them onto curved and irregular aerospace components encountered in aircraft \cite{khan2024systematic, chuang2021multi}. 
The slip casting method can be employed to fabricate the ceramic-based open cell structures like SA, which involves pouring a ceramic slurry into a porous mould, where capillary action forms a solid layer that is later sintered \cite{tallon2010effect}. 
Complex geometries can also be fabricated using the Ultrafast Shaping and Sintering (USS) method \cite{shan2024programmable}, which utilises a programmable Joule heater to thermally activate ceramic green compacts, making them into shapes like auxetics before rapid sintering. 
For additive approaches, ceramic 3D printing \cite{zhang2022additive} builds components layer-by-layer from a ceramic-polymer feedstock, which requires subsequent debinding and sintering. Chmielewski \cite{chmielewski2016metal} has reported a gradient fabrication approach to join ceramic and metal layers. The sharp interface between ceramic and metal layers has been eliminated by the gradient approach, which guarantees a smooth transition from ceramic to metal. The above literature confirms the possibility of fabricating SA-based BTO structure of the nozzle.


The RL performance of the Radar Absorbing Structures (RAS) is often evaluated using computationally intensive techniques such as Finite Difference Frequency Domain (FDFD) or Finite Difference Time Domain (FDTD), Method of Moments (MOM) and Finite Element Methods (FEM) \cite{ayari2024advanced}. However, analysing composite-based metamaterials for EM absorption characteristics, which contain a heterogeneous mixture of nanoparticles, using the above-mentioned numerical techniques can be very challenging and computationally expensive. The methodology employed in this work (i.e., variational asymptotic method-based homogenisation and the modified transfer matrix (MTM) method) is one of the efficient approaches to analyse complex unit cell architectures or composite structures.
The homogenisation method entails substituting an equivalent homogenous block of material for the complex geometry \cite{xu2025size, khan2025homogenized}, while taking into account the effects of the original structure without making any performance-deterrent assumptions to make the computation simpler.
Combining material homogenisation and the physics-based tool (MTM method), which uses scattering matrices, can be reliable for the proposed mechanical metamaterials-based EM absorbers.
The scattering matrices evaluated in the MTM tool are dependent on the thickness, along with the homogenised EM properties \cite{rumpf2011improved} of the auxetic barium titanate nanoparticles based metamaterial, which results in identifying the influence of these dependent parameters on the absorption characteristics of the auxetic structure.

In the current study, three key novelty contributions are conceptually appealing and practically relevant to developing advanced radar-absorbing structures.
First, the RL characteristics of auxetic metamaterials made from barium titanate, a ceramic-based material, are evaluated, thereby providing a detailed understanding of how the intrinsic EM properties of ceramics, when designed as metamaterials (auxetic structures), influence microwave absorption in the 8-18 GHz frequency band.
Second, the research investigates the significant weight-saving benefits of open-cell structures such as the proposed SA architecture. Comparative studies prove that SA geometry is beneficial in reducing material usage, thereby reducing structural weight and providing better absorption characteristics than traditional monolithic PB ceramic absorbers in many cases.
Third, this work bridges the gap between academic research and real-world application by implementing the SA designs into a nozzle assembly and carrying out a thorough examination of RCS reduction; hence, confirming the practical advantage of auxetic metamaterial based absorbers compared to the monolithic ceramic blocks. 
Additionally, the proposed methodologies and design principles used in this paper are easily implementable in other frequency bands and alternative platform systems, further establishing the work's generality and significance in developing next-generation, lightweight, and high-performance electromagnetic signature management solutions. 
The article is organised into four sections to discuss the research work. Section \ref{sec:Methodology} discusses the in-house tools and the Floquet port analysis carried out to verify the in-house tool, along with describing the process employed to evaluate the RCS.
In the Section \ref{sec:validation}, the in-house tools are validated with experimental and numerical analysis results. Subsequently, in Section \ref{sec:thicknessandmassadvantage}, the RL performance of the SA and PB are compared for different BTO material configurations to evaluate the SA architectures that can provide the dual advantage of mass saving and better absorption capabilities.
Later in Section \ref{sec:RCSevaluation} selected SA structures from the previous analysis have been used to determine the rear aspect RCS performance of aircraft nozzle (with and without BTO).
Later, the inferences drawn from the studies carried out are discussed in Section \ref{sec:conclusions}.

\section{Materials details and methodology}
\label{sec:Methodology}
\begin{figure}[!ht]
    \centering
    \includegraphics[width=01.0\linewidth]{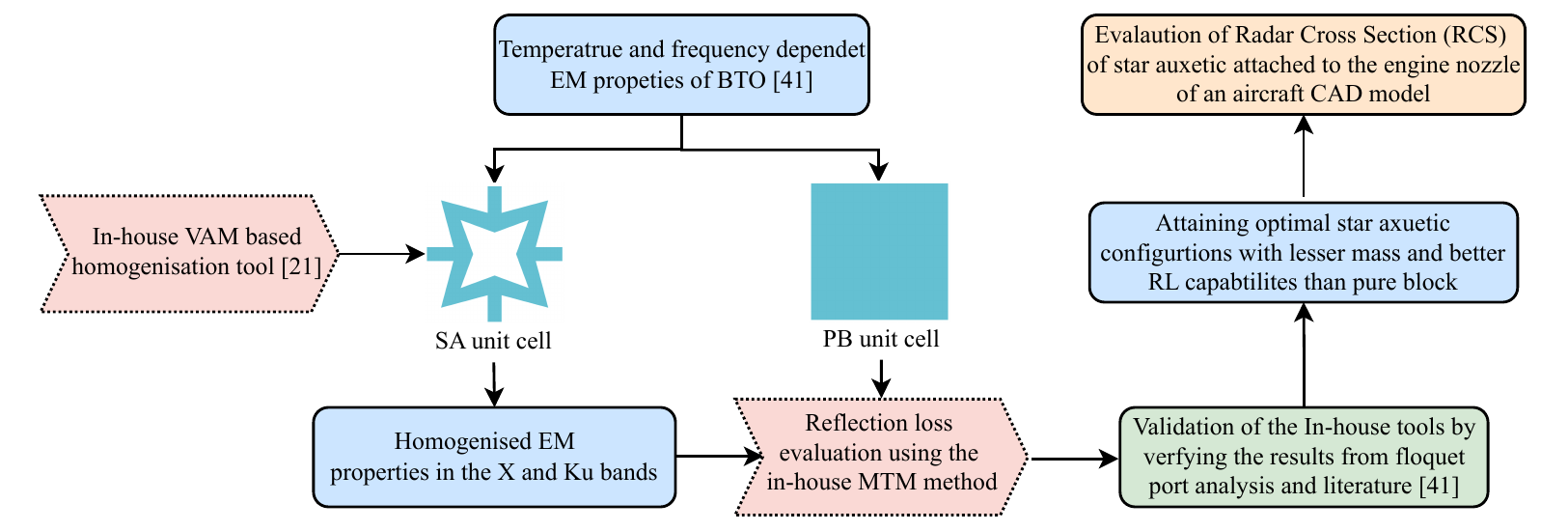}
    \caption{Schematic representing the flowchart of the methodology being implemented for evaluating the RCS of an aircraft when a star auxetic structure is attached to the engine nozzle.} 
    \label{fig:flowchart}
\end{figure}

To present a comprehensive basis for further analyses, two key performance measures used in this study are outlined: (i) RL measures the fraction of incident radar energy that is absorbed by a material and is determined in decibels (dB); the lower the RL value, the greater the effectiveness of the absorption. (ii) RCS measures the strength of the scattering of the electromagnetic waves towards the source radar and aids in determining how well a target is viewed. Lower RCS is linked with better stealth capabilities, which makes the target less visible to radar. In the proposed research work, the RL spectra are obtained for the BTO based SA and PB configurations in the X and Ku bands.
Star-based auxetic is the only mechanical metamaterial that is considered in this work based on the inferences reported by Kumar et al. \cite{kumar2023computational}.
The temperature and frequency-dependent properties of the BTO assigned to the SA are considered from the experimental results reported by Saini et al. \cite{saini2016dual}.
Both the SA and PB are placed in the free space with a metal backing to evaluate the RL capabilities.
The proposed work aims to understand the advantages of using a SA instead of a PB for EM absorption applications.
The parametric studies are carried out to evaluate the advantages of using an auxetic metamaterial by comparing the RL of the SA and PB: (a) same thickness, (b) same weight and (c) similar RL spectrum. Some of the results from parametric studies (a) are described in the subsequent section; the other results from (a), and the entire results from (b) and (c) are carried out but not shown here for the sake of brevity.
Subsequently, studies are carried out to identify SA configurations that can provide similar or enhanced RL, bandwidth, and weight savings compared with the PB.
Once the SA samples that can provide better RL are identified, a few of those configurations are attached to a nozzle of an open-source aircraft CAD model \cite{sutrakar2014effect} to evaluate the RCS capabilities of SA in comparison with PB, thereby assessing the advantage for practical applications.
Figure \ref{fig:flowchart} shows the flowchart depicting the proposed work overview. A detailed explanation of the methodologies and modelling strategies implemented for carrying out the analysis is given in the subsections below.
\subsection{Materials details}
\begin{figure}[!ht]
    \centering
      \subfloat{{\includegraphics[width=0.5\textwidth]{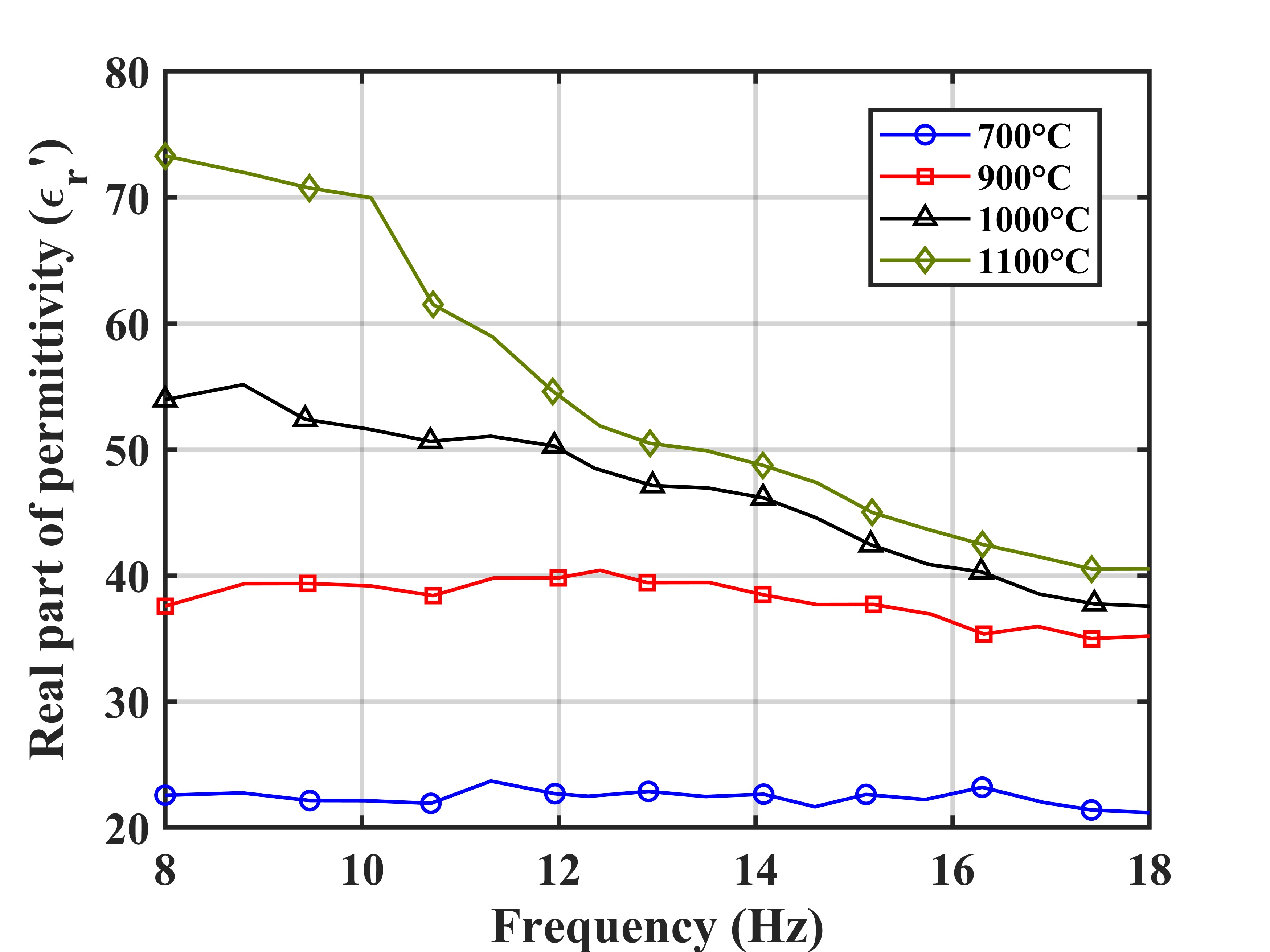} \label{sfig: SEMmultiphasecompositeimages1} }}
  \subfloat{{\includegraphics[width=0.5\linewidth]{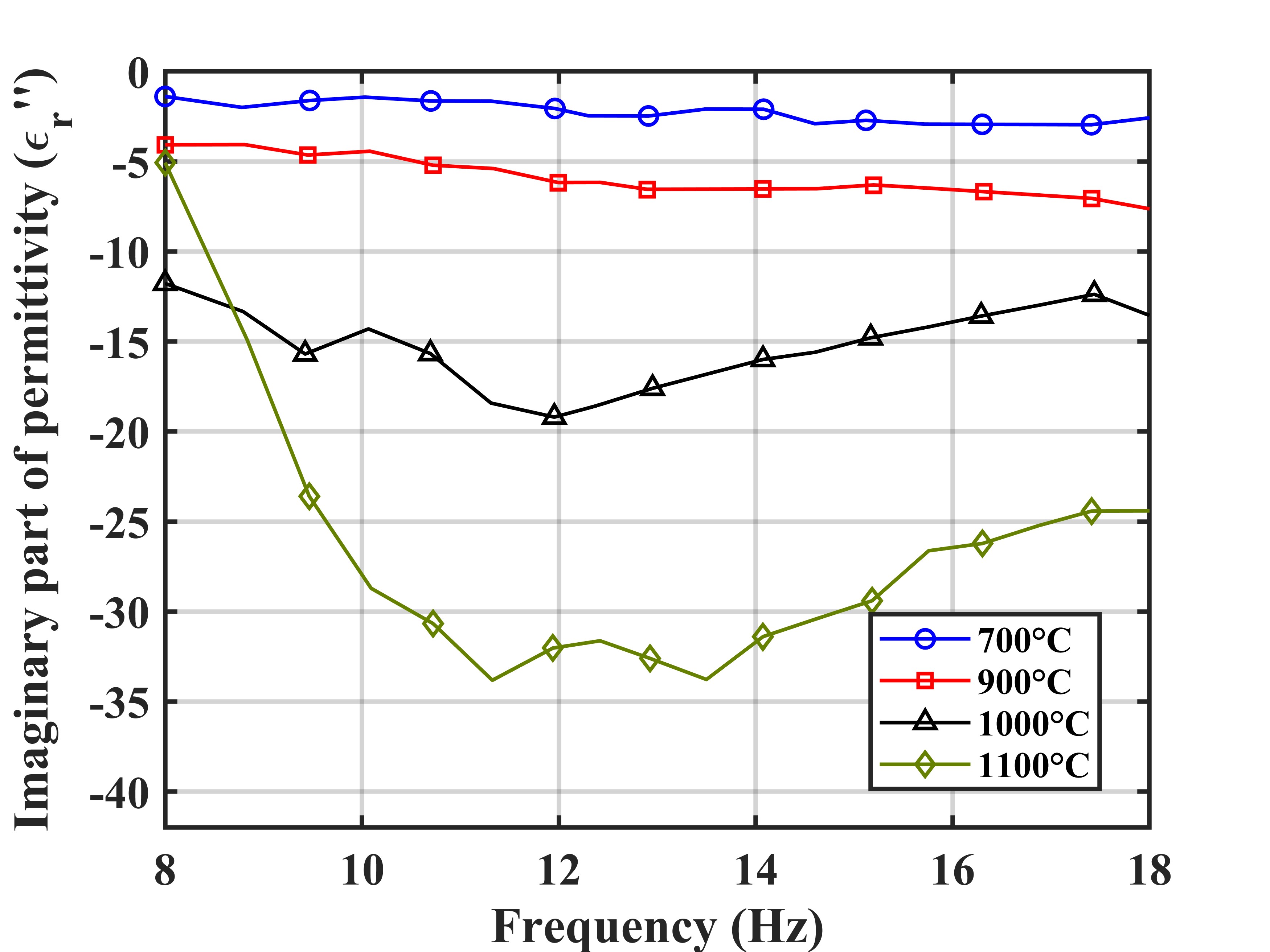} \label{sfig: SEMmultiphasecompositeimages2}}}
  \captionsetup{labelsep=period}
  \caption{Real and imaginary parts of the relative electric permittivity of annealed BTO in the X and Ku frequency band \cite{saini2016dual}.}
  \label{fig:annealedBTOproperties}
\end{figure}

Electromagnetic data based on BTO nanoparticles annealed at 700$^{\circ}$C (BTO700), 900$^{\circ}$C (BTO900), 1000$^{\circ}$C (BTO1000), and 1100$^{\circ}$C (BTO1100) are used. 
Annealing BTO nanoparticles results in sequential phase development, increasing tetragonality in BaTiO\({}_{3}\) and thereby influencing the EM absorption capabilities. 
Lower temperatures ($\leq$ 700$^{\circ}$C) result in incomplete crystallisation, while higher temperatures promote the displacement of Ti\({}^{4+}\) ions within the TiO\({}_{6}\) octahedra, which is crucial for enhancing ferroelectric polarisation and dielectric loss for improved EM absorption \cite{saini2016dual}. Electromagnetic characterisation over the 8–18 GHz band showed that the complex permittivity of BTO is strongly influenced by annealing temperature. The real part of permittivity ($\epsilon_\text{r}^{'}$) increased with temperature, while the imaginary part ($\epsilon_\text{r}^{''}$) exhibited significant variation, with the 1100$^{\circ}$C sample showing the highest loss values. These changes are attributed to dipole relaxation mechanisms arising from the ferroelectric nature of tetragonal BTO. The observed increase in dielectric loss with temperature shows the role of thermal treatment in tailoring the microwave absorption performance of BTO-based materials. For reference, the trend in complex permittivity as a function of annealing temperature reported by Saini et al. \cite{saini2016dual} is summarised in Figure \ref{fig:annealedBTOproperties}, which shows the variation of both real and imaginary electrical permittivity across the 8-18 GHz frequency range.
\subsection{Electromagnetic properties homogenisation}
\label{sec:vamemhomogenisation}
\textcolor{black}{This section briefly describes the formulation used in developing the in-house tools \cite{kumar2023computational}. The proposed approach of homogenising the complex structure and evaluating the RL capabilities using the in-house tools results in a computationally efficient method to carry out RL studies.
RL capabilities of the SA depend on both the frequency and temperature-dependent homogenised electromagnetic properties. The proposed Variational Asymptotic Method (VAM) based homogenisation formulation is used in the present study to attain the effective electromagnetic properties of auxetics.
The material response of the barium titanate (BTO) is governed by the combined electromagnetic constitutive equations as,}
\begin{equation}
    \begin{bmatrix}
        \textbf{D}\\
        \textbf{B}
    \end{bmatrix} = 
    \begin{bmatrix}
        \bm{\epsilon} & 0\\
        0 & \bm{\mu}
    \end{bmatrix} \begin{bmatrix}
        \textbf{E}\\
        \textbf{H}
    \end{bmatrix}
\label{eq:electromagneticconstitutiveequations}
\end{equation}
\textcolor{black}{Where $\bm{\epsilon}$ is the dielectric permittivity tensor, $\bm{\mu}$ is the magnetic permeability tensor, $\textbf{E}$ is electric field vector, $\textbf{H}$ is magnetic field vector, $\textbf{D}$ is electric displacement vector, and $\textbf{B}$ is magnetic flux density vector.
The framework relies on two fundamental assumptions: first, that the macroscopic electromagnetic properties are unaffected by boundary conditions, and second, that the global electric and magnetic potential fields are equivalent to the volume-averaged local fields. The second assumption can be ensured by choosing the centre of the auxetic unit cell as the origin. Then the local fields can be written as, }
\begin{equation}
\begin{aligned}
        \Phi_{e} & = \dfrac{1}{V}\int_{V}\phi_{e}\text{dV} \;\;;\;\; \phi_{e} = \Phi_{e} + \varphi_{e}\\
         \Phi_{m} & = \dfrac{1}{V}\int_{V}\phi_{m}\text{dV} \;\;;\;\; \phi_{m} = \Phi_{m} + \varphi_m
\end{aligned}
\label{eq:secondassumption}
\end{equation}
\textcolor{black}{Where $\varphi_{e}$ is electric fluctuation field, $\varphi_{m}$ is magnetic fluctuation field, $\Phi_{e}$ is global electric potential field, $\Phi_{m}$ is global magnetic potential field, $\phi_{e}$ is local electric potential field, $\phi_{m}$ is local magnetic potential field, and V represents the volume of the entire structure.}

\begin{figure}[!ht]
    \centering
    \includegraphics[width=0.35\linewidth]{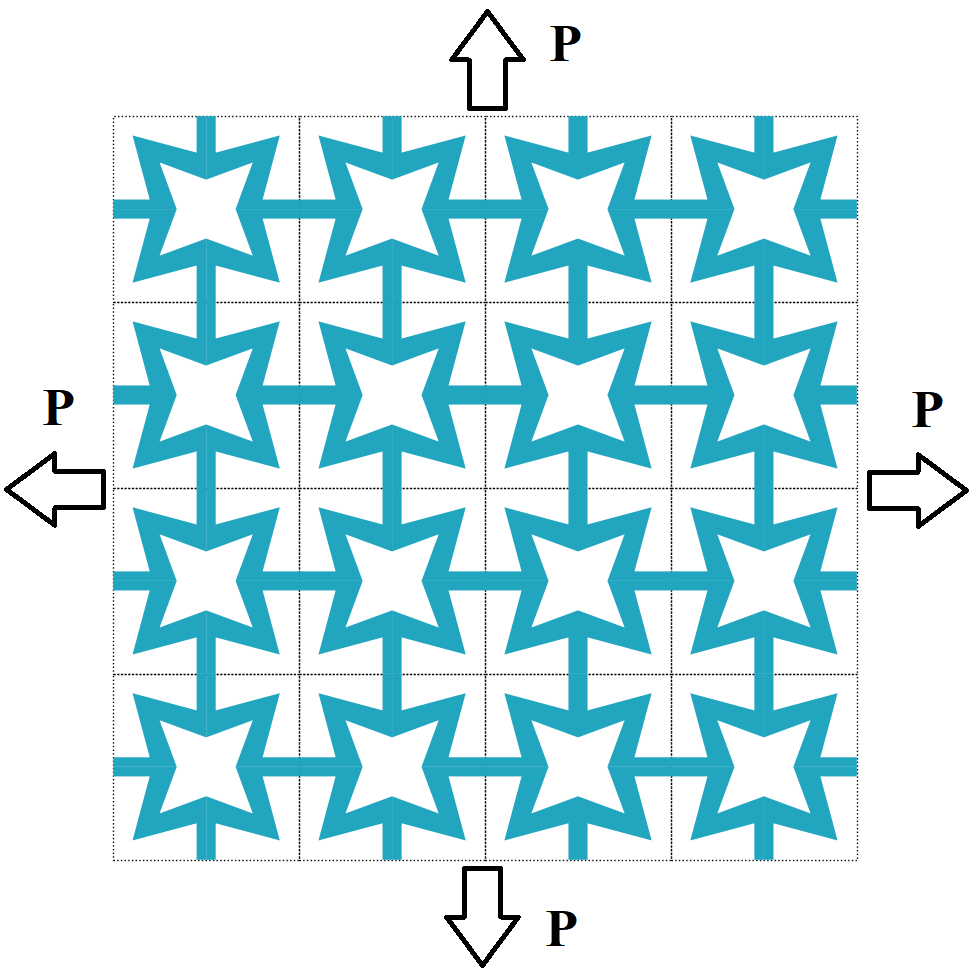}
    \caption{Schematic depicting the 2D representation of the periodic star-based auxetic unit cell in x and y directions.}
    \label{fig:starpatternrepetition}
\end{figure}

\textcolor{black}{The two-dimensional repetitive unit cell (RUC) of the SA is shown in Figure \ref{fig:starpatternrepetition}. Periodicity of the fluctuation fields ($\varphi_{e}^{+x} = \varphi_{e}^{-x}$, $\varphi_{e}^{+y} = \varphi_{e}^{-y}$) in the x and y directions are used to generate the whole structure with RUCs, where $\varphi_{e}^{+x}$ represents electric fluctuation field in positive x-direction, $\varphi_{e}^{-x}$ represents electric fluctuation field in negative x-direction, $\varphi_{e}^{+y}$ represents electric fluctuation field in positive y-direction, and $\varphi_{e}^{-y}$ represents electric fluctuation field in negative y-direction.}
\textcolor{black}{The effective electromagnetic properties of this SA structure are attained by minimising the energy functional ($J_{\text{auxetic}}$) of the SA structure. The functional can be reduced to just the electromagnetic energy component ($J_{\text{auxetic}} = \langle \Pi \rangle$) by enforcing that the RUC's centre is at the origin and also the periodicity of the RUC \cite{tang2008variational}. 
The resulting energy functional can be obtained as,}
\begin{equation}
\begin{split}
    \langle \Pi \rangle &= \dfrac{1}{\Omega} \int_{\Omega}  \begin{bmatrix}
        \textbf{E}\\
        \textbf{H}
    \end{bmatrix}^{\text{T}} \begin{bmatrix}
        \bm{\epsilon} & 0\\
        0 & \bm{\mu}
    \end{bmatrix}  \begin{bmatrix}
        \textbf{E}\\
        \textbf{H}
    \end{bmatrix} d\Omega 
\end{split}
\label{eq:resultingenergy}
\end{equation}
\textcolor{black}{Where $\langle \Pi \rangle$ is the volume-averaged electromagnetic energy density, $\Omega$ is the volume of a single RUC, and the superscript $\text{T}$ denotes the transpose operation.
$J_{\text{auxetic}}$ needs to be minimised to obtain the effective electromagnetic properties. After minimisation and solving using finite element analysis, the effective electromagnetic properties are obtained as,}
\begin{equation}
\overline{\textbf{M}} = \begin{bmatrix}
        \bm{\overline{\epsilon}} & 0\\
        0 & \bm{\overline{\mu}}
    \end{bmatrix}  = \bm{M_{lg}}^T \bm{M_{ll}}^{-1} \bm{M_{lg}} + \bm{M_{gg}}
    \label{eq:feabasedhomogenousproperteistenrosr}
\end{equation}
\textcolor{black}{Where $\overline{\textbf{M}}$ is the effective homogenised material property tensor, $\bm{\overline{\epsilon}}$ is the effective dielectric permittivity tensor, $\bm{\overline{\mu}}$ is the effective magnetic permeability tensor. More detialed explanation and derivation of individual terms can be referred from \cite{kumar2023computational, kumar2024novel}.
The geometric modelling and the element discretisation of the SA are carried out using an open-source computational tool, Gmsh \cite{gmsh}. The in-house homogenisation tool to attain effective EM properties is developed in Julia \cite{Julia-Language}, integrated with the libraries of Gmsh \cite{gmsh}.} After mesh convergence studies, a finite element model of the auxetic unit cell containing $\approx$ $1.50 \times 10^5$ a number of elements, containing a large proportion of hexahedral elements and a small number of tetrahedron elements, has been used to evaluate the homogenised properties.

Input temperature-dependent electromagnetic properties of the BTO in the X and Ku bands are obtained from the experimental results reported by Saini et al. \cite{saini2016dual}. The temperature-dependent effective EM properties are obtained at every 0.05 GHz frequency interval. 

\subsection{Reflection loss evaluation using MTM}
The homogenised EM properties obtained from Section \ref{sec:vamemhomogenisation} are given as input to the in-house modified transfer matrix (MTM) tool to evaluate the RL spectra of the SA structure in the 8 GHz to 18 GHz frequency.
The MTM formulation starts with obtaining the electromagnetic fields inside the homogeneous layer by solving Maxwell's equations. The equations are normalised with ${\bm{H^{'}}} = -j\sqrt{\mu_0/\epsilon_0}\bm{H}$ to attain a robust numerical solution.
\begin{equation} \label{eq: Maxwell-equations}
    \begin{aligned}
        \bm{\nabla} \times \bm{H^{'}} &= K_{0}\bm{{\epsilon}^\text{T}}(f)\bm{{E}} \\
        \bm{\nabla} \times \bm{E} &= K_{0}\bm{{\mu}^\text{T}}(f)\bm{{H^{'}}}
    \end{aligned}
\end{equation}
\textcolor{black}{Where $\bm{H^{'}}$ is the normalised magnetic field vector, $j$ is the imaginary unit ($\sqrt{-1}$), $\mu_0$ is the permeability of free space, $\epsilon_0$ is the permittivity of free space, $K_0$ is the free space wave number, $\bm{{\epsilon}^\text{T}}(f)$ is the temperature and frequency dependent electrical permittivity tensor, and $\bm{{\mu}^\text{T}}(f)$ is the temperature and frequency dependent magnetic permeability tensor.
The assumptions in the MTM method are that the EM wave is a plane wave and the material is homogeneous. Based on these two assumptions, the partial differential equations reduce to four ODEs. The four equations are combined to form an eigenvalue problem as, }
\begin{equation}\label{eq: elec-mag-eigen}
    \dfrac{d^{2}}{d{z^{'}}^{2}}\begin{bmatrix}
        E_{x}\\
        E_{y}
    \end{bmatrix} - RS\begin{bmatrix}
        E_{x}\\
        E_{y}
    \end{bmatrix} = \begin{bmatrix}
        0\\
        0
    \end{bmatrix}
\end{equation}
\textcolor{black}{Where $E_{x}$ is the x-component of electric field, $E_{y}$ is the y-component of electric field, ${z^{'}}$ is the normalised spatial coordinate in z-direction (${z^{'}} = K_0 z$), $R$ is the first coefficient matrix from the ODE system, $S$ is the second coefficient matrix from the ODE system, and $RS$ is the product matrix of R and S.
Solving the Eigenvalue problem provides the EM field solution $\bf{\Psi}$$^{T}$ with superscript T representing the annealing temperature. After getting the electromagnetic field components, the EM field continuity is ensured at the boundaries of the structure and the free space. The boundary conditions that need to be satisfied are, }
\begin{equation} \label{eq: field-bcs}
    \begin{aligned}
         \bm{\Psi}^T({z^{'}} = 0) &= \bm{\Psi}_{1} \\
         \bm{\Psi}^T({z^{'}} = K_0 d) &= \bm{\Psi}_{2}
    \end{aligned}
\end{equation}
\textcolor{black}{Where $\bm{\Psi}_{1}$ is the electromagnetic field vector at the first interface (z=0), $\bm{\Psi}_{2}$ is the electromagnetic field vector at the second interface (z=d), $\bm{\Psi}^T$ is the temperature-dependent electromagnetic field solution within the material, $d$ is the thickness of the homogenised layer, and $K_0 d$ is the normalised thickness.}
\begin{figure}[!ht]
    \centering
    \includegraphics[width=1.0\linewidth]{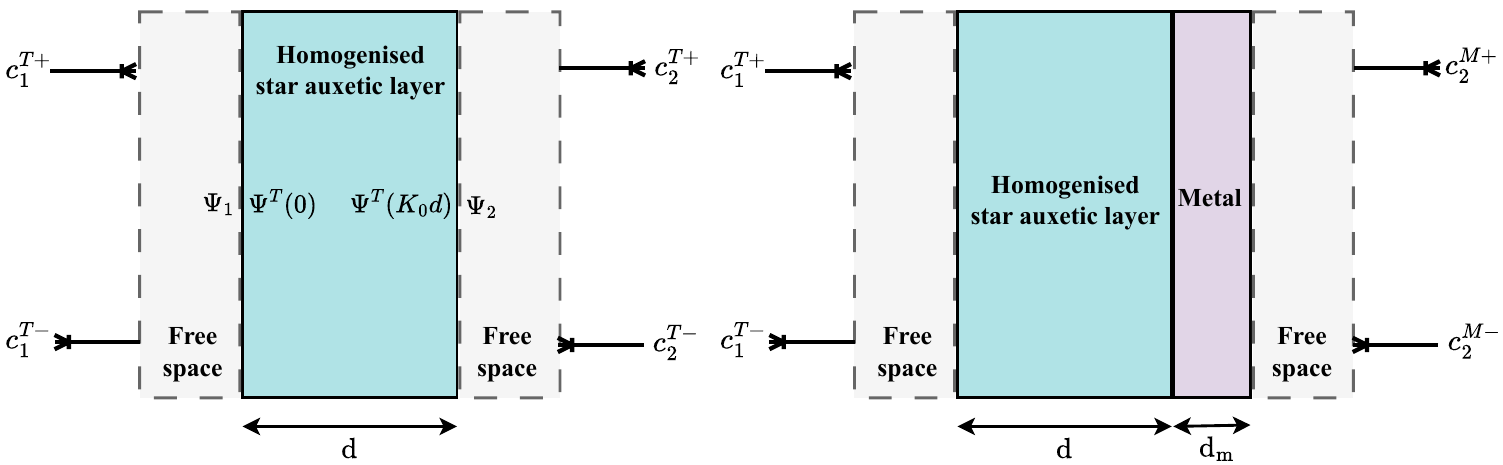}
    \caption{Schematic represents the homogenised star auxetic layer of thickness d, with the free space medium on both sides of the homogeneous layer. The electromagnetic continuity conditions ($\bm{\Psi}$) at the interfaces of the layers are also represented appropriately. The absorber with metal backing, considered in the analysis for evaluating the RL spectrum, is also represented.}
    \label{fig:representationimage}
\end{figure}
\textcolor{black}{After applying the boundary conditions, the field coefficients are rearranged to relate the transmission and reflected coefficients with the incident coefficient using the scattering parameters. Using both the scattering matrices for the auxetic layer and metal backing, the global scattering matrix is evaluated using Redheffer's star product \cite{WANG2020235}. }
\begin{equation}\label{eq: scattering-matrix2}
    \begin{bmatrix}
        c_{\text{ref}} \\ c_{\text{trans}}
    \end{bmatrix}
    = \text{\bf{S}}^{G} \begin{bmatrix} c_{\text{inc}} \\ 0 \end{bmatrix}
\end{equation}
\textcolor{black}{Where $c_{\text{ref}}$ is the reflected field coefficient vector, $c_{\text{trans}}$ is the transmitted field coefficient vector, $c_{\text{inc}}$ is the incident field coefficient vector, and $\text{\bf{S}}^{G}$ is the global scattering matrix.
The electric field components in the reflection and transmission regions are obtained using the amplitude coefficients. Later, the reflection component can be obtained, and from there, the RL spectrum for all the BTO material configurations is attained for the SA structure. The more detailed explanation to attain the equations can be found in \cite{kumar2025integrated}.}
\subsection{Floquet port analysis}
\label{sec:Floquet port analysis}
A PB and star-shaped auxetic (SA) structure made up of Barium Titanate (BTO) is modelled with PEC as back reflecting plate as a reference unit cell (refer to Figure \ref{fig:representationimage}) to investigate its electromagnetic behaviour under periodic excitation across the 8-18 GHz frequency range. The intrinsic temperature-dependent material properties of BTO have been obtained from \cite{saini2016dual}. The main focus of the present study is to generate the reflection coefficient (S$_{11}$) of the BTO infinite array using a unit cell with periodic boundary conditions. Inorder to replicate these conditions, Floquet ports are assigned to the top face of the unit cell to launch and collect the electromagnetic wave while accounting for higher-order diffraction modes that can occur in
periodic structures \cite{sakyi2024waveguide}. The lateral faces are defined using lattice pair (periodic) boundary conditions, ensuring that the tangential components of the electric and magnetic fields match perfectly from one side to the other. This setup effectively models an infinite two-dimensional repetition of the unit cell,
eliminating unwanted edge effects and providing an accurate representation of practical periodic configurations. The reflection coefficient (S$_{11}$) is extracted to quantify the ratio of reflected energy and incident energy by the structure. Fresnel has been generated for the SA and PB for different thicknesses, varying the incident angles. The S$_{11}$ results obtained from the PB and SA simulations have been used to validate the in-house tools. The S$_{11}$ generated using the Floquet port analysis in Ansys \cite{Ansys} is utilised directly for the RCS analysis instead of modelling the entire 3D SA structure on the nozzle of the aircraft. 

\subsection{RCS analysis}
\begin{figure}[!ht]
    \centering
      \subfloat{{\includegraphics[width=0.5\textwidth]{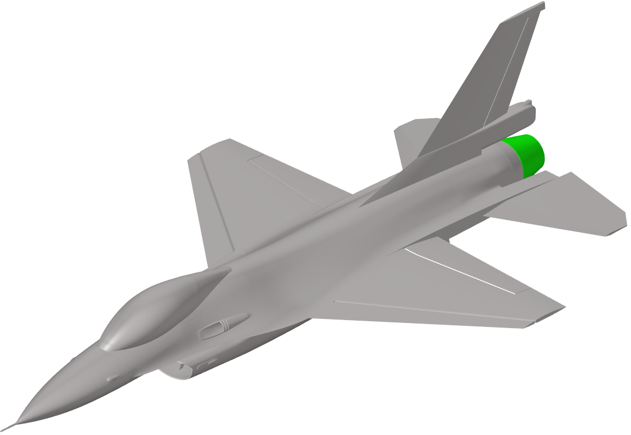} \label{sfig: CAD_Model_1} }}
  \subfloat{{\includegraphics[width=0.5\linewidth]{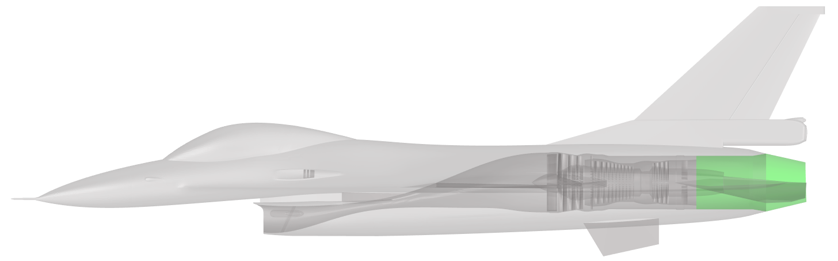} \label{sfig: CAD_Model_2}}}
  \captionsetup{labelsep=period}
  \caption{Open source CAD model of a fighter aircraft: (a) Isometric view, and (b) Side view. Nozzle is highlighted in Green colour.}
  \label{fig:cadmodels}
\end{figure}

An open-source aircraft CAD model \cite{sutrakar2014effect} is considered for evaluating the BTO performance at the nozzle, as shown in Figure \ref{fig:cadmodels}. The performance of the aircraft nozzle with BTO is evaluated, and the results are compared with the perfect electrical conducting (PEC) nozzle.  The generated Fresnel from the Floquet port analysis has been assigned to the nozzle regions of the aircraft model. Later, high-frequency EM simulations are carried out using shooting and bouncing rays with 20 bounces and 5 ray densities \cite{sutrakar2014effect}. RCS simulations are carried out from 8 to 18 GHz of frequencies with a step size of 1GHz for both horizontal (HH) and vertical (VV) polarisations, for $\theta$ varying from 70 degrees to 110 degrees and $\phi$ varying from -60 degrees to +60 degrees with a step of 0.5 degrees. The angles are chosen such that it aids in covering the complete rear aspect of the aircraft in azimuthal and elevation regions. The rear aspect of the aircraft corresponds to $\theta$=90 degrees and $\phi$=0 degrees. The median RCS at a given frequency is also calculated with varying $\theta$ and $\phi$ to evaluate the rear sector performance of the aircraft holistically.

RCS simulations of aircraft with a PEC material are initially carried out as a reference. Subsequently, two different configurations of BTO are considered in the nozzle region, i.e. (a) PB and (b) SA. Four different BTO material configurations are considered, where BTO is annealed at 700$^{\circ}$C, 900$^{\circ}$C, 1000$^{\circ}$C, and 1100$^{\circ}$C at the material level to obtain better RL performance. Also, an aircraft nozzle with varying BTO thicknesses is considered for comparing the results of PB and SA. 

\section{Results and Discussion}
\label{sec:resultsanddiscussion}

\subsection{Validation of in-house Homogenisation-MTM computational tools with experimental and Floquet port analysis}
\label{sec:validation}
In this section, the validation of the in-house tools is carried out 
in two stages: (a) validation of the MTM tool and (b) validation of the combined tool (Homogenisation and MTM). 
\subsubsection{Validation of the MTM tool}
\begin{figure}[!ht]
    \centering
      \subfloat{{\includegraphics[width=0.5\textwidth]{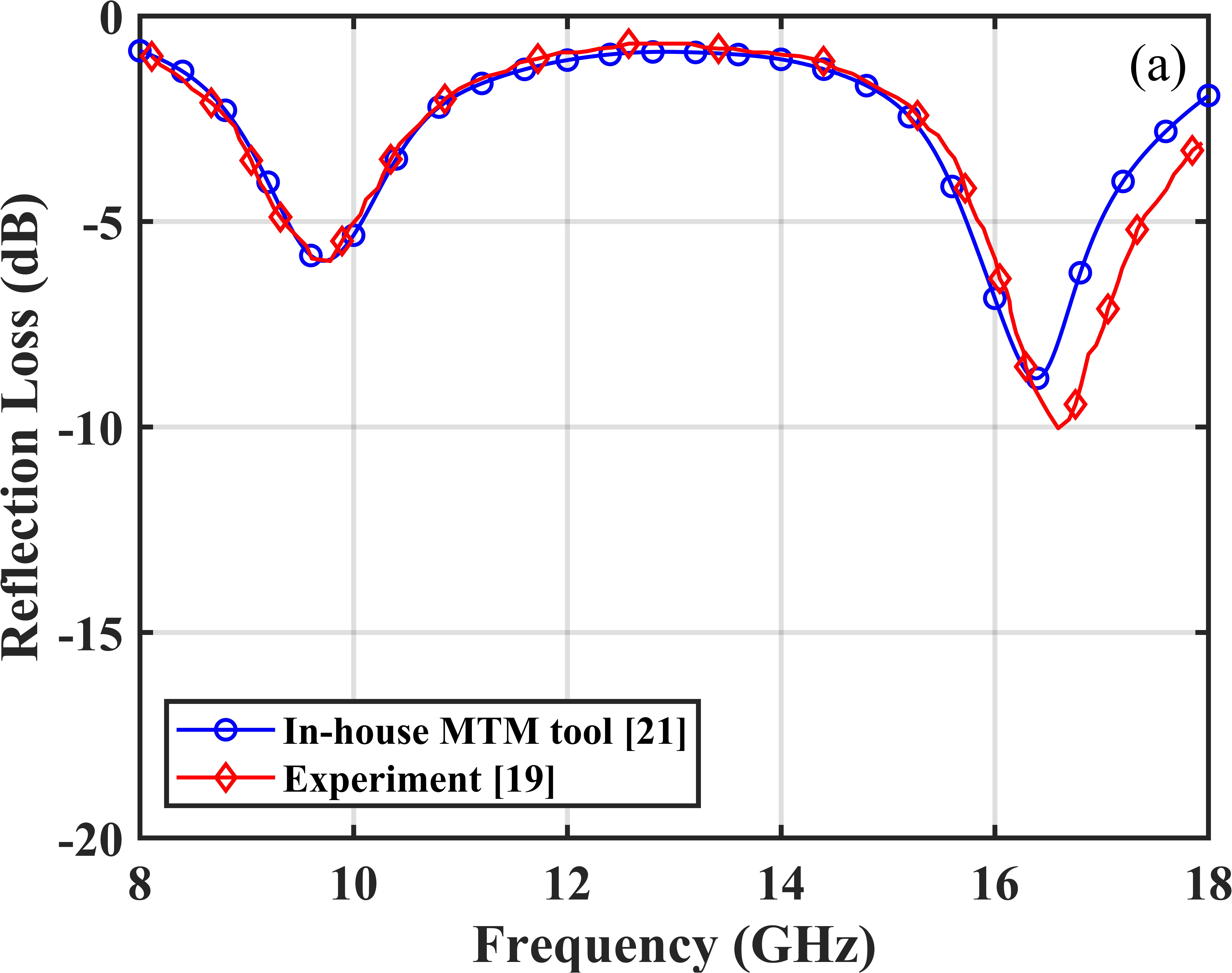} \label{sfig: SEMmultiphasecompositeimages1} }}
  \subfloat{{\includegraphics[width=0.5\linewidth]{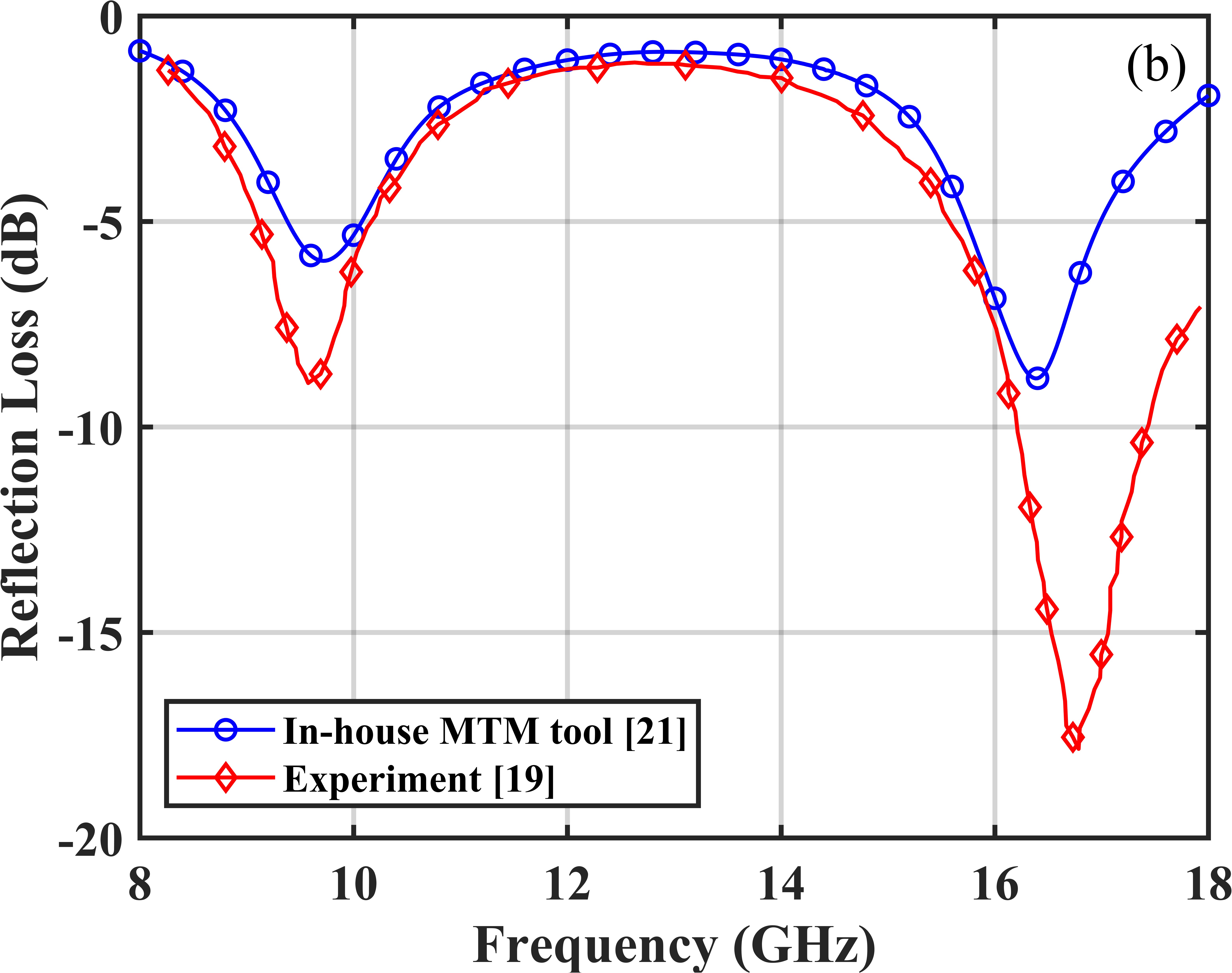} \label{sfig: SEMmultiphasecompositeimages2}}}
  \captionsetup{labelsep=period}
  \caption{RL spectra in the X and Ku bands for the rubber-based composite loaded with annealed BTO nanoparticles (annealed till 1100$^{\circ}$C): (a) 8.0 mm thick rubber-based composite with 70\% weight of BTO nanoparticles and (b) 6.5 mm thick rubber-based composite with 80\% weight of BTO nanoparticles.}
  \label{fig: validationwithreferencepaperresults}
\end{figure}

To validate the MTM tool, the RL spectra from the MTM tool are compared with the experimentally reported RL by Saini et al. \cite{saini2016dual}, as shown in Figure \ref{fig: validationwithreferencepaperresults}.
The close matching of the RL spectra obtained from the MTM tool and experiments \cite{saini2016dual} confirms the validity of the in-house MTM tool.

\subsubsection{Validation of the combined tool (Homogenisation and MTM)}

\begin{figure}[!ht]
    \centering
      \subfloat{{\includegraphics[width=0.5\textwidth]{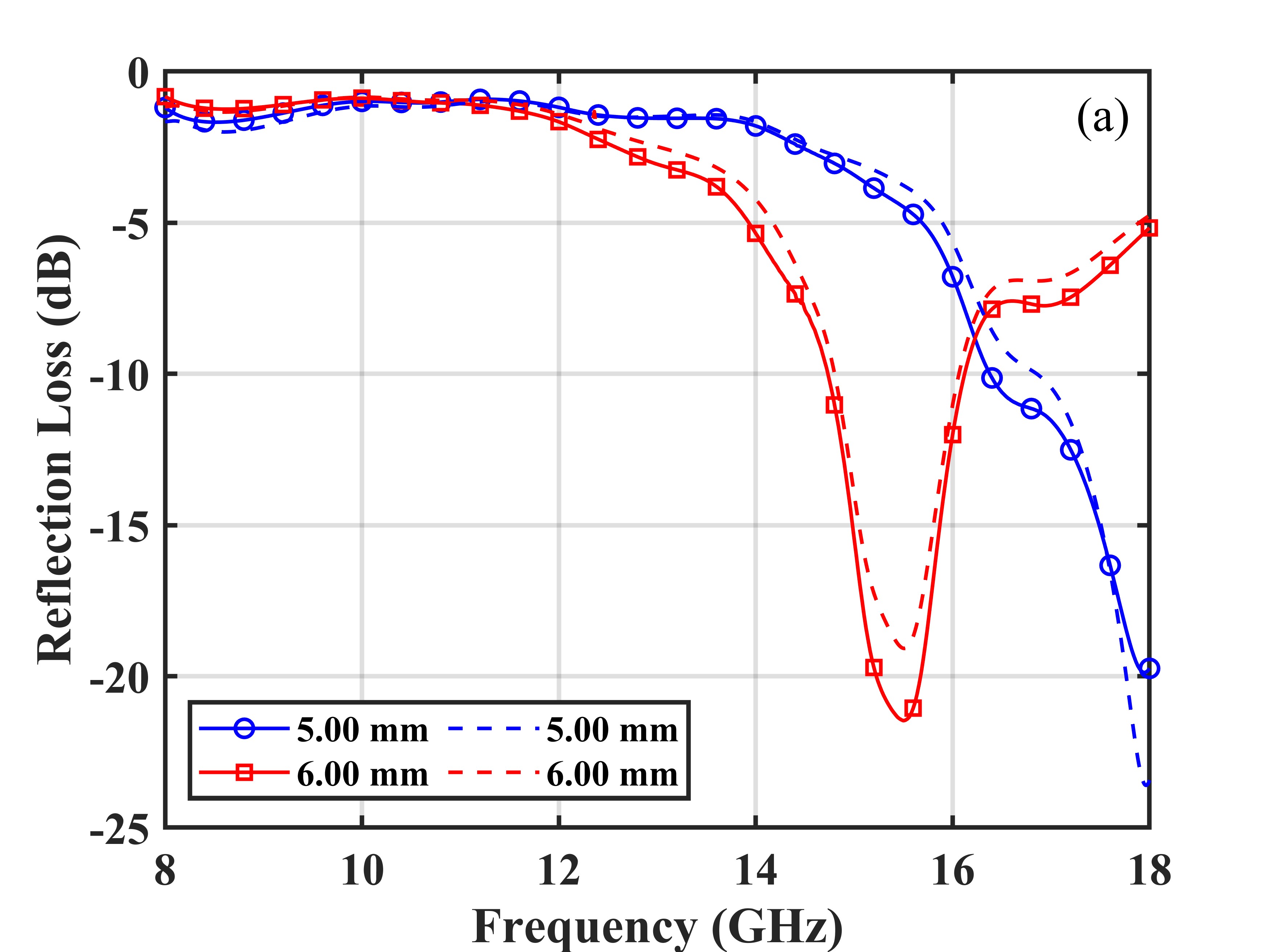}\label{fig:BTO700validation} }}
  \subfloat{{\includegraphics[width=0.5\linewidth]{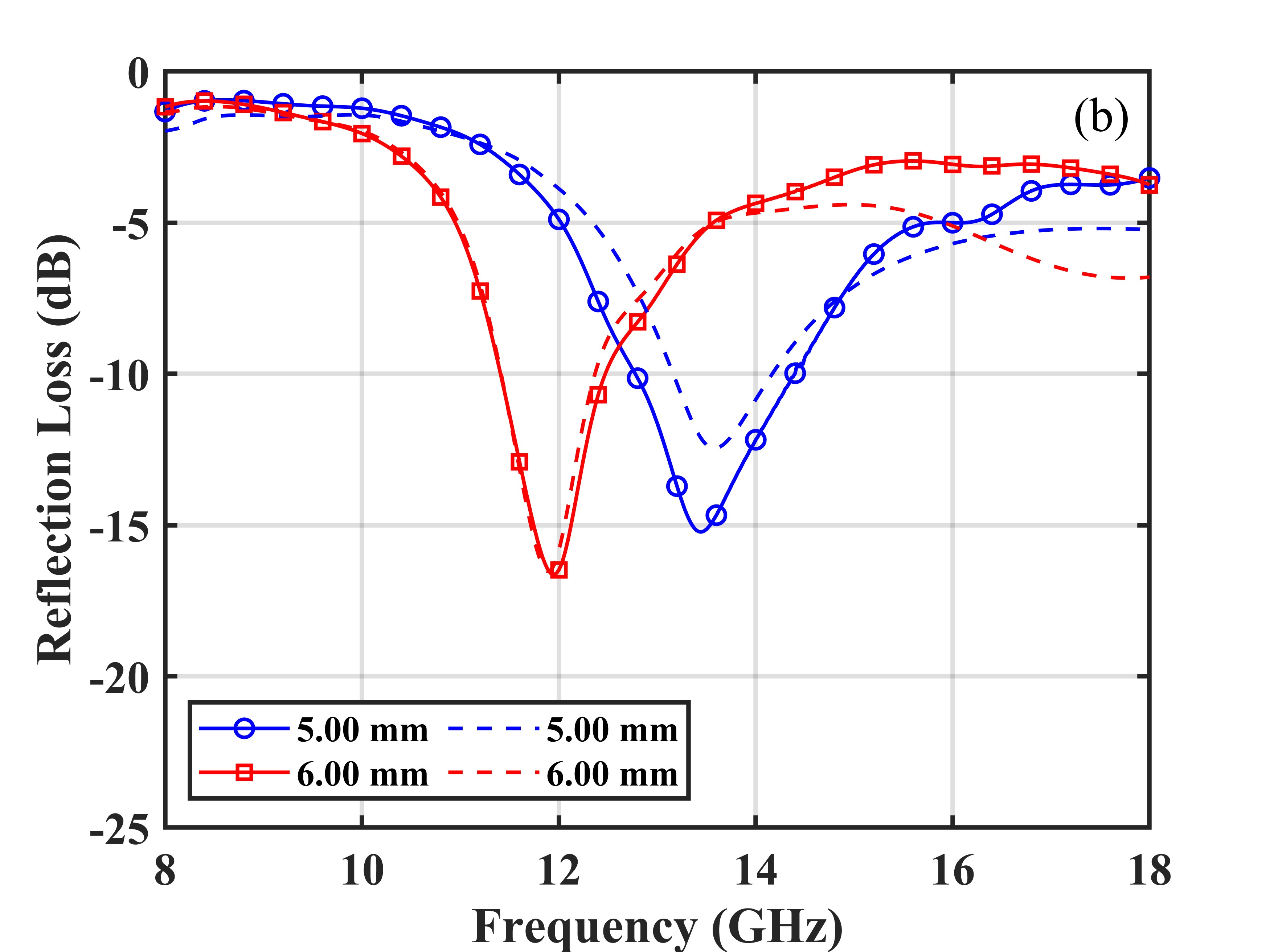}\label{fig:BTO900validation}}}\\ \vspace{-0.35cm}
        \subfloat{{\includegraphics[width=0.5\textwidth]{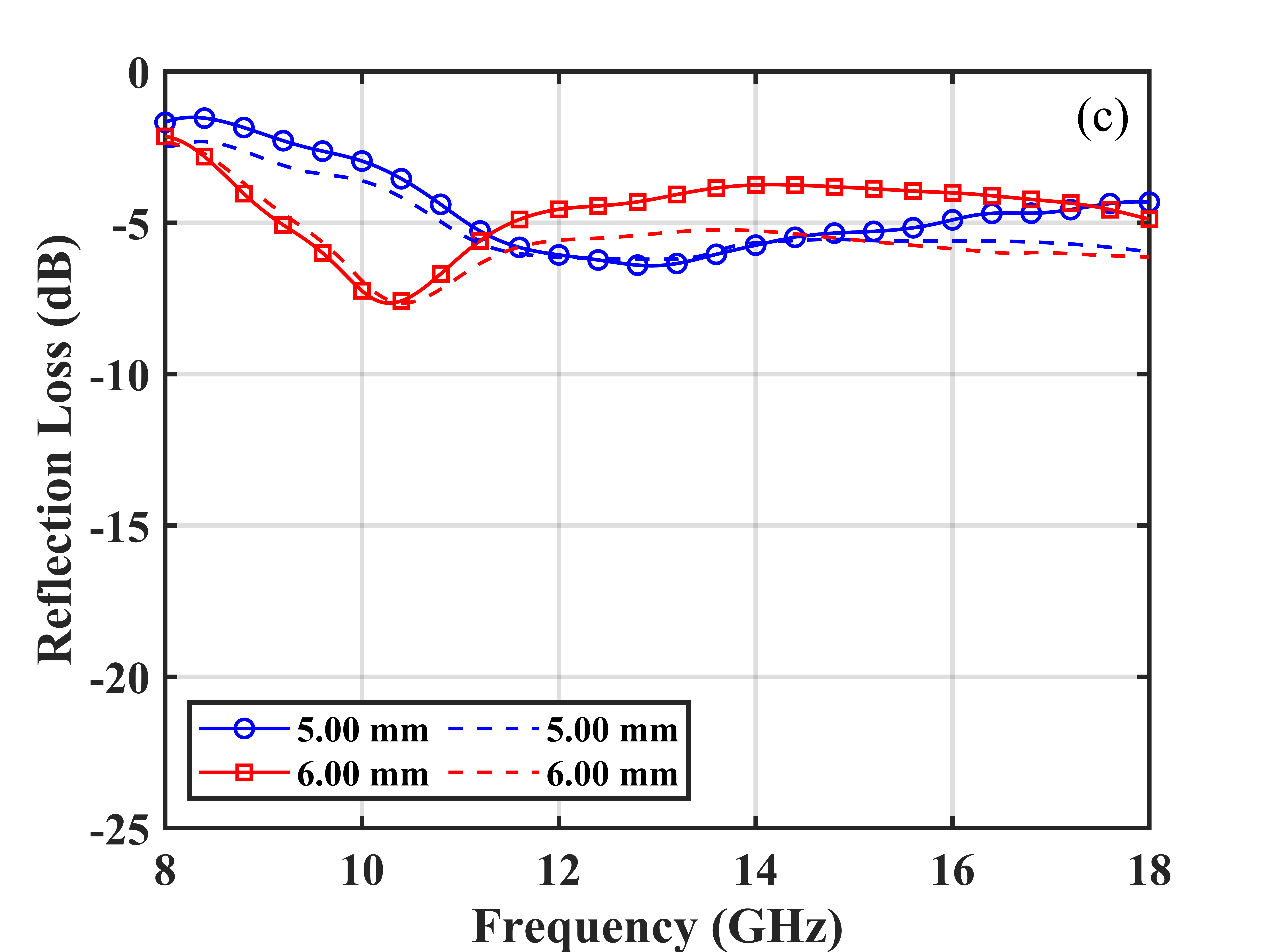}\label{fig:BTO1000validation} }}
  \subfloat{{\includegraphics[width=0.5\linewidth]{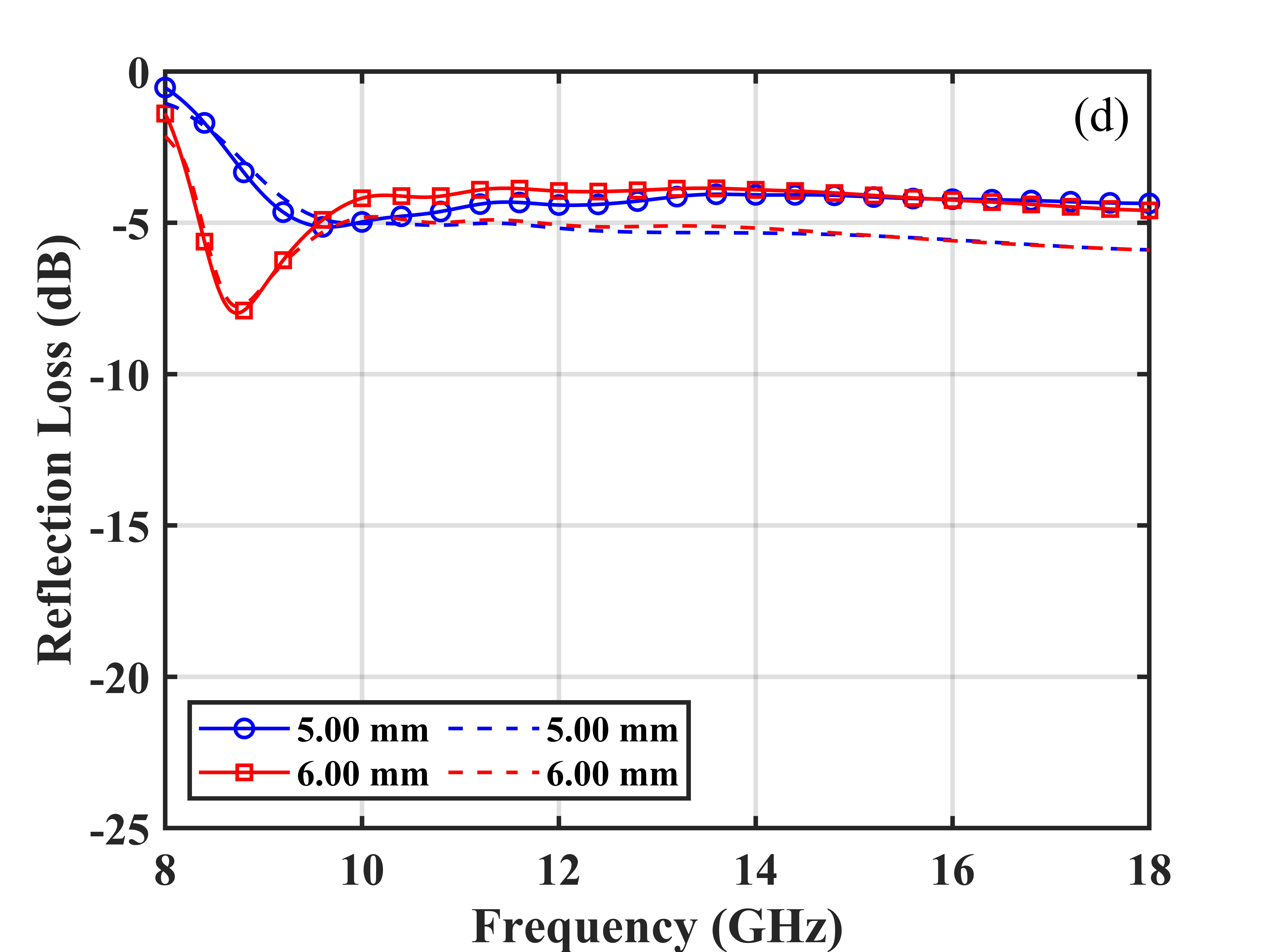}\label{fig:BTO1100validation}}}
  \captionsetup{labelsep=period}
  \caption{RL spectra obtained from homogenization-MTM methodology and Floquet port analysis of star-auxetic samples of 5.00 mm and 6.00 mm thick in the X and Ku frequency bands. Star samples are made up of (a) BTO nanoparticles annealed at 700$^{\circ}$C,  (b) BTO nanoparticles annealed at 900$^{\circ}$C,  (c) BTO nanoparticles annealed at 1000$^{\circ}$C,  (d) BTO nanoparticles annealed at 1100$^{\circ}$C. The solid and dashed lines in the plots represent the results obtained from the in-house tools and Floquet port analysis, respectively.}
\label{fig:RLforvalidationofBTOsamples}
\end{figure}

In order to validate the homogenisation-MTM tool, Floquet port analysis is performed on the SA configurations as described in Section \ref{sec:Floquet port analysis}.
Two random thicknesses of 5.0 mm and 6.0 mm are considered for comparing RL spectra.
Figure \ref{fig:RLforvalidationofBTOsamples} shows good agreement between the MTM (solid lines) and Floquet port (dashed lines) results. 
Subsequent parametric studies are carried out using the validated tool, varying SA and PB thicknesses from 1.00 to 10.00 mm in 0.05 mm steps, giving 181 configurations per type and 32,761 (181 $\times$ 181) SA–PB combinations.
Representative results are discussed in the following sections.

Beyond RL evaluation, the proposed combined tool offers substantial computational advantages. Both the proposed methodology and Floquet port analysis are carried out on a system equipped with an Intel i7-7700 K CPU, 32 GB of RAM, and a Windows operating system. The Floquet port analysis of a 6.00 mm thick BTO900 material configuration required 8.62 minutes, while the proposed combined tool completed the same analysis in only 3.18 minutes, with approximately 63\% reduction in computational time. 

\subsection{Comparison of Reflection loss spectra of BTO-based samples}
\label{sec:thicknessandmassadvantage}
\subsubsection{Reflection loss spectra of flat PB and SA samples with the same thickness}
This section aims to provide a systematic comparison of the RL behaviour of the BTO-based flat panel samples modelled as PB and SA geometries.
The study is carried out to understand the influence of sample thickness on the EM absorption capabilities in terms of attaining RL $<$-5dB (considered as a benchmark value for evaluating the bandwidth, which is defined as the continuous frequency range in which RL $<$ -5 dB) in the 8-18 GHz frequency range.
The investigations on the RL spectra on different thickness samples (PB and SA samples) of different BTO configurations will allow to identify the variation in the spectra of the samples.
Even though the sample thicknesses are increased from 1.00 mm to 10.00 mm and the comparison studies are carried out, only a few cases are shown in this section; the remaining RL comparison plots are not shown here for the sake of brevity. 



\begin{figure}[!ht]
    \centering
      \subfloat{{\includegraphics[width=0.5\textwidth]{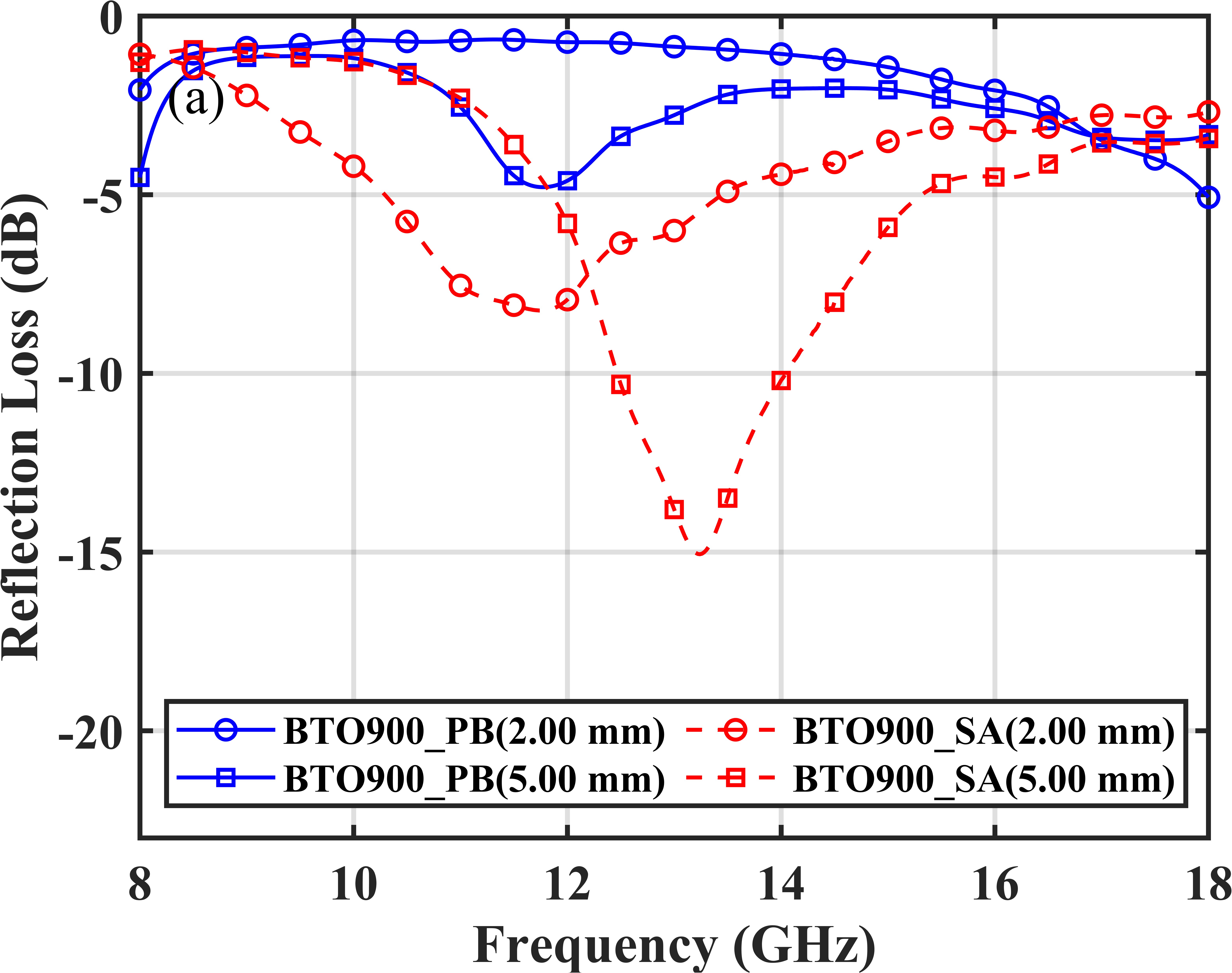}\label{fig:BTO9001mmand2mm} }}
        \subfloat{{\includegraphics[width=0.5\textwidth]{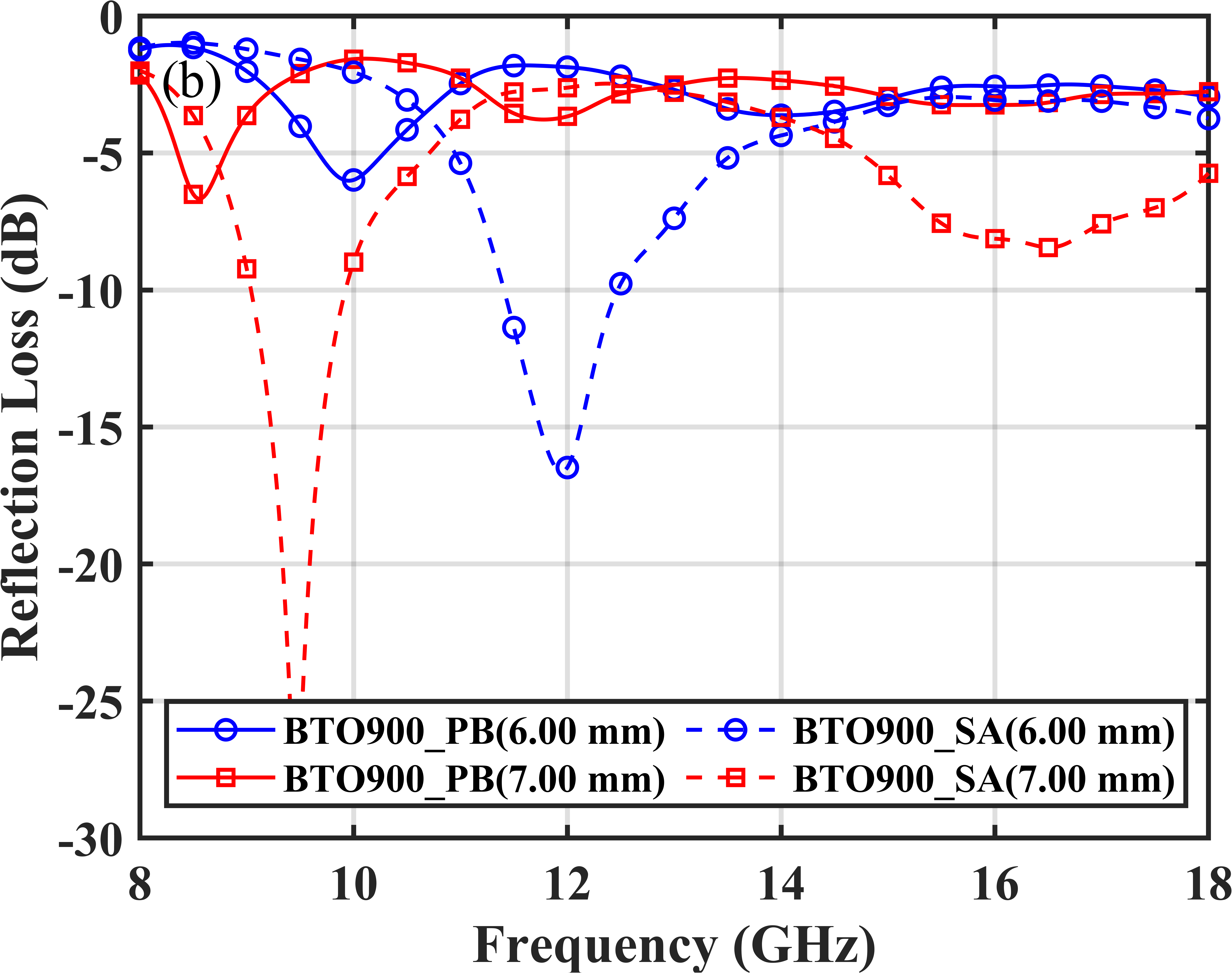}\label{fig:BTO9006mmand7mm} }}
  \captionsetup{labelsep=period}
  \caption{RL spectra of the Pure block and Star auxetic samples made up of BTO900 (Barium Titanate annealed till 900$^{\circ}$C) with the same thickness. (a) 2.00 mm and 5.00 mm thick, (b) 6.00 mm and 7.00 mm thick.}
\label{fig:RLofBTO900bothpureblockandstarauxetic}
\end{figure}

\textcolor{black}{Figure \ref{fig:RLofBTO900bothpureblockandstarauxetic} provides the RL spectra comparison of PB and SA samples made up of BTO900. For the 2.00 mm and 5.00 mm thick samples (Figure \ref{fig:BTO9001mmand2mm}), the 5.00 mm SA sample shows high EM absorption, with a minimum RL of -15.00 dB. The 2 mm and 5 mm SA samples have a combined bandwidth of 3.15 GHz and 3.61 GHz, respectively. As the thickness increases, the SA samples of 6.00 mm and 7.00 mm thickness show a notable improvement in RL performance (Figure \ref{fig:BTO9006mmand7mm}). The 6.00 mm-thick SA sample has a minimum RL of -16.57 dB and a bandwidth of 02.64 GHz. The 7.00 mm-thick SA sample achieves a minimum RL of -27.00 dB and a bandwidth of 05.32 GHz, showing better absorption than the corresponding PB samples. These results show that the auxetic architecture enables enhanced absorption at mid-thickness levels for BTO900 samples. Among all the considered PB and SA samples, most of the SA samples have shown a balance between absorption characteristics and bandwidths in both X and Ku bands.}

\textcolor{black}{These results point out the fact that both the annealing temperature during the material synthesis and geometry have a significant influence on attaining the minimum RL and bandwidth. 
The absorption performance is highly thickness-dependent, with SA samples providing a clear advantage in many cases.
To distinguish the EM absorption capabilities for BTO900 configuration in the X and Ku bands, the minimum RL and bandwidth are reported in Table \ref{table:PB_SA_900_X_Ku_Band}. 
In the X-band, SA samples outperform PB, showing minimum RL $<$-10 dB for thicknesses between 5 mm and 10 mm, with bandwidths $\geq$ 1.00 GHz.
In the Ku-band, all SA samples except those with 1.00 mm and 3.00 mm thicknesses achieve the required RL, while PB samples only meet this at 1.00 mm and 3.00 mm thicknesses. The details for other material configurations of BTO700, BTO1000 and BTO1100 have also shown the advantage of using SA design, but not shown here for the sake of brevity.}

\begin{table}[!ht]
\centering
\caption{Comparison of minimum RL and bandwidth of PB and SA samples with varying thickness (1.00 mm to 10.00 mm) made up of material configuration BTO900 in the X-band and Ku-band frequency ranges.}
\resizebox{1.00\textwidth}{!}{\begin{tabular}{ccccccccccc}
\hline
\multirow{2}{*}{Thickness (mm)} & \multirow{2}{*}{Band} &  \multicolumn{2}{c}{RL (dB)} & \multicolumn{2}{c}{BW (GHz)} & \multirow{2}{*}{Band} & \multicolumn{2}{c}{RL (dB)} & \multicolumn{2}{c}{BW (GHz)} \\ \cline{3-6} \cline{8-11}
 & & PB & SA & PB & SA & & PB & SA & PB & SA \\ \hline
   1 & \multirow{10}{*}{X-band} & -23.13 & -00.44 & 1.21 & 0.00 & \multirow{10}{*}{Ku-band} & -22.82 & -02.77 & 2.00 & 0.00 \\
   2 & & -02.07 & -08.24 & 0.00 & 1.72 & & -05.07 & -07.94 & 0.00 & 1.43 \\
   3 & & -07.60 & -06.93 & 0.77 & 0.33 & & -07.26 & -02.42 & 0.45 & 0.00 \\
   4 & & -10.67 & -02.32 & 0.98 & 0.00 & & -04.60 & -13.20 & 0.00 & 2.87 \\
   5 & & -04.79 & -04.89 & 0.00 & 0.00 & & -04.61 & -15.22 & 0.00 & 3.61 \\
   6 & & -06.01 & -16.63 & 0.54 & 1.05 & & -03.62 & -16.48 & 0.00 & 1.59 \\
   7 & & -06.67 & -28.72 & 0.46 & 2.05 & & -03.67 & -08.45 & 0.00 & 3.27 \\
   8 & & -04.88 & -29.29 & 0.00 & 1.26 & & -03.20 & -08.82 & 0.00 & 2.92 \\
   9 & & -04.45 & -11.01 & 0.00 & 0.97 & & -03.23 & -09.21 & 0.00 & 2.64 \\
  10 & & -04.97 & -10.43 & 0.00 & 1.65 & & -03.00 & -06.91 & 0.00 & 3.31 \\ \hline
\end{tabular}}
\label{table:PB_SA_900_X_Ku_Band}
\end{table}

\begin{figure}[!ht]
    \centering
      \includegraphics[width=0.85\textwidth]{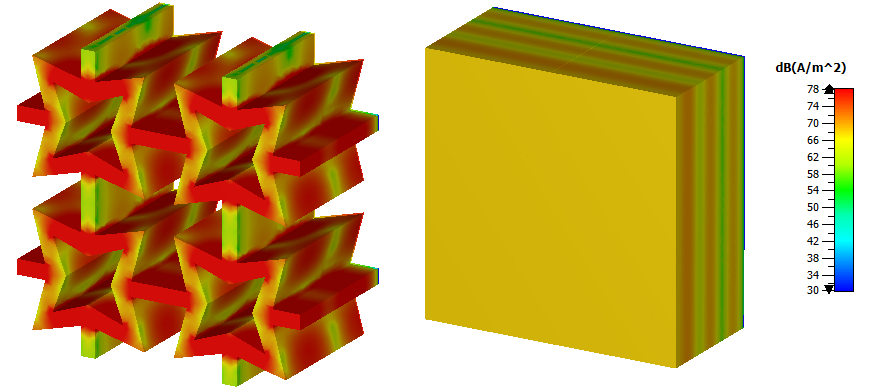}
  \captionsetup{labelsep=period}
  \caption{Current density distribution (in dB(A/m$^{2}$)) at 12 GHz for 6.00 mm thick BTO900 based SA and PB structures.}
\label{fig:currentdensitystarPB}
\end{figure}

The difference in electromagnetic absorption between the SA and the PB can be clearly understood from the absolute current density ($\lvert$J$\rvert$) distribution. To identify the difference in this distribution, the current density variation of 6.00 mm thick BTO900-based unit cells is shown in Figure \ref{fig:currentdensitystarPB}.
In the PB structure, the induced current density is relatively weak and almost uniformly spread over the large flat top surface, with values being lower (below $\approx$ 66 dB(A/m$^{2}$)). This uniform and low-intensity current (compared to SA) results in poor absorption of the solid block.
In contrast, the SA structure shows a different current density distribution. The current density is strongly concentrated along the narrow arms, especially at the sharp tips of each star, reaching very high values, as indicated by the bright red regions (above 75-78 dB(A/m$^{2}$)). These intense current density hotspots are present throughout the unit cell. Since the power dissipated inside the absorber is proportional to the square of the current density, these localised high current density regions drastically increase the losses and, therefore, enhance the RL capabilities.
This highly non-uniform and enhanced current density distribution in the SA structure is therefore the main reason for its superior absorption performance compared to the conventional PB structure.

\subsubsection{Reflection loss spectra of flat PB and SA samples with enhanced RL and bandwidth}
\label{sec:betterRLPBSA}
\begin{figure}[!ht]
    \centering
      \subfloat{{\includegraphics[width=0.5\textwidth]{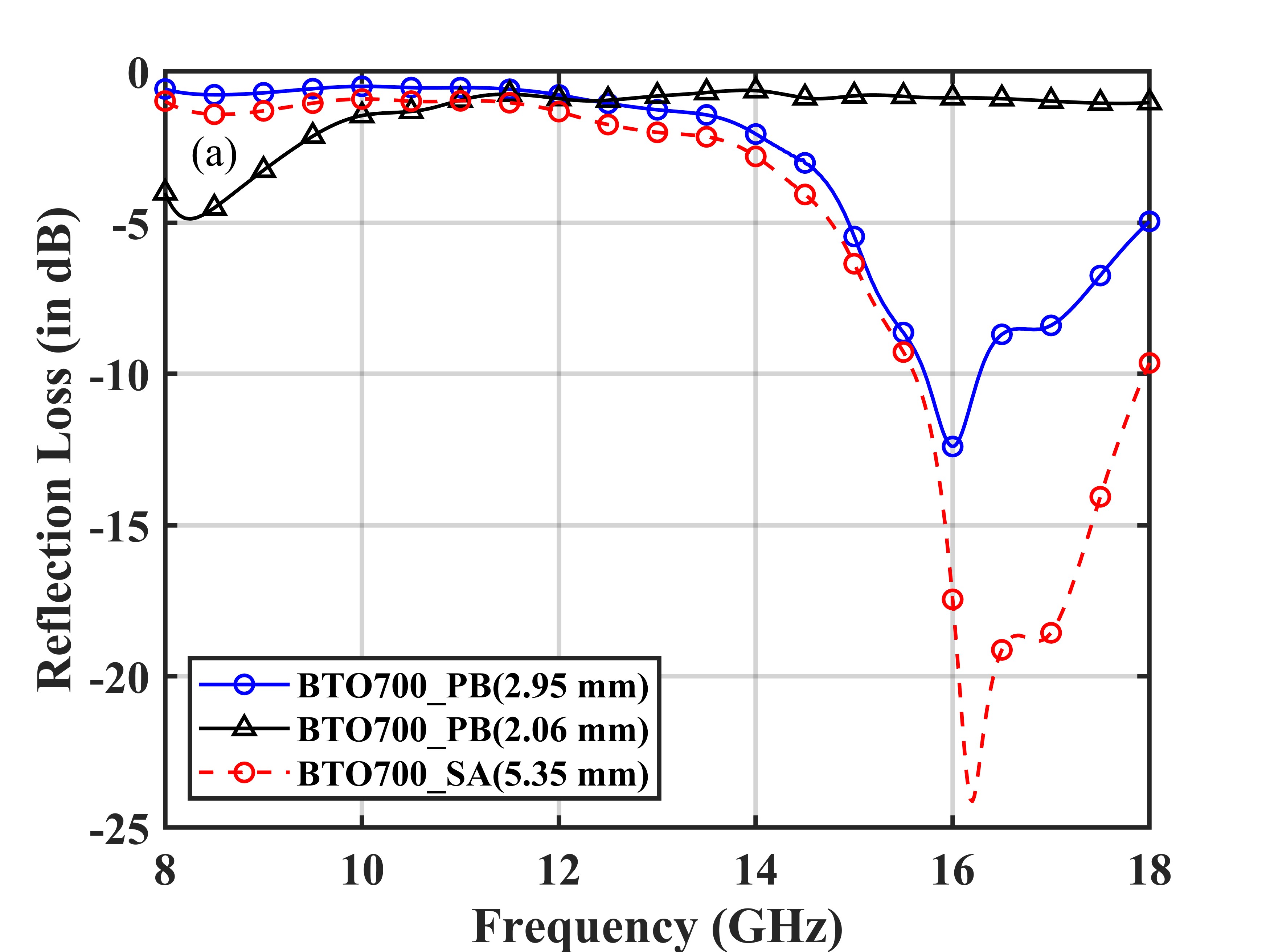}\label{fig:betterRLBTO7001} }}
      \subfloat{{\includegraphics[width=0.5\textwidth]{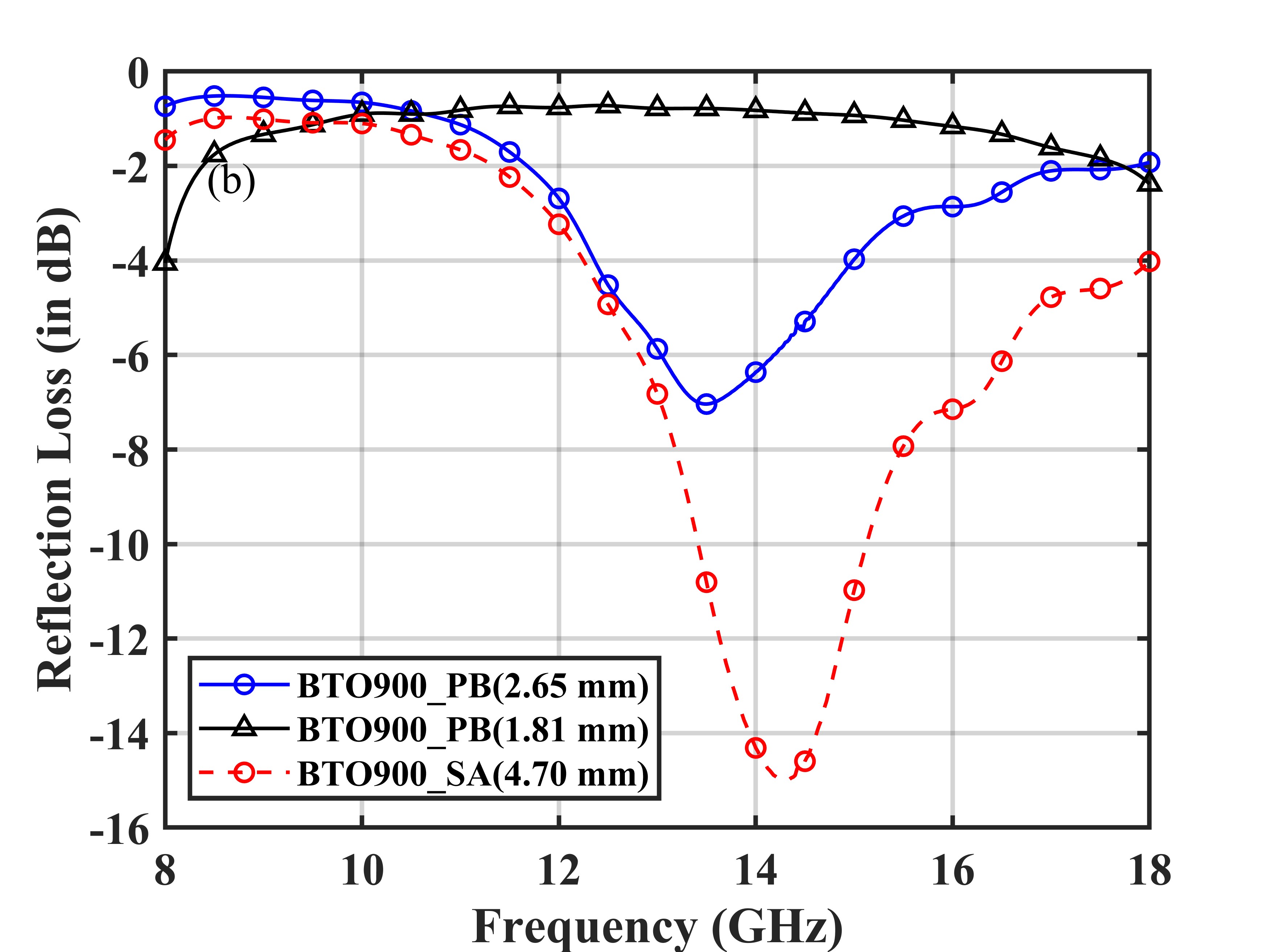}\label{fig:betterRLBTO9001} }}
  \captionsetup{labelsep=period}
  \caption{Comparison of the RL spectra of Pure block and Star auxetic samples made up of BTO700 (Barium Titanate annealed till 700$^{\circ}$C) and BTO900 (Barium Titanate annealed till 900$^{\circ}$C), where Star auxetic samples exhibited better RL capabilities. (a) BTO700: 2.95 mm and 5.35 mm thick, (b) BTO900: 2.65 mm and 4.70 mm thick. The plots in black colour represent the RL spectra of PB samples that have the same weight as the SA samples.}
\label{fig:betterRLBTO700}
\end{figure}

\begin{figure}[!ht]
    \centering
      \subfloat{{\includegraphics[width=0.5\textwidth]{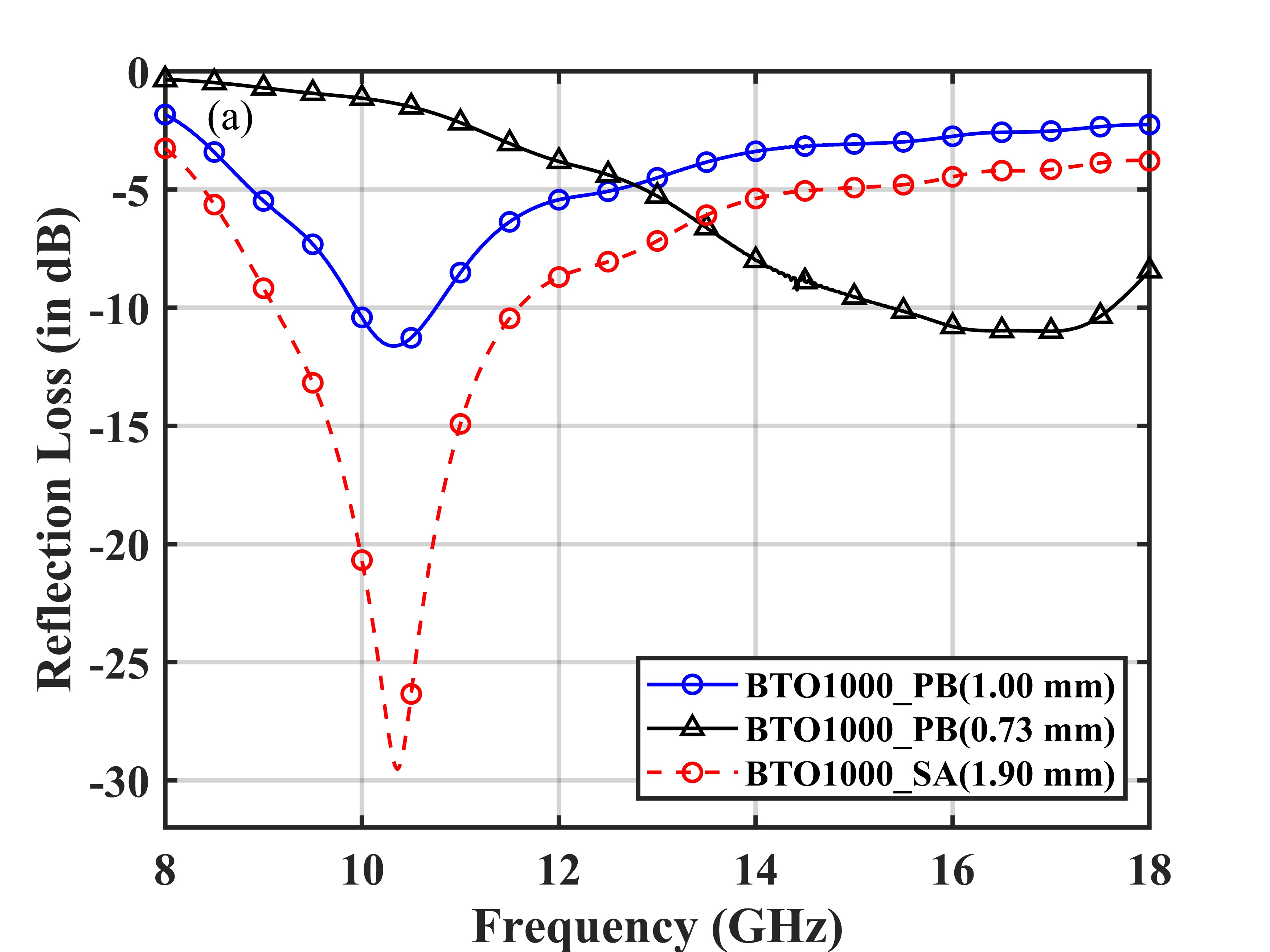}\label{fig:betterRLBTO1000_1} }}
        \subfloat{{\includegraphics[width=0.5\textwidth]{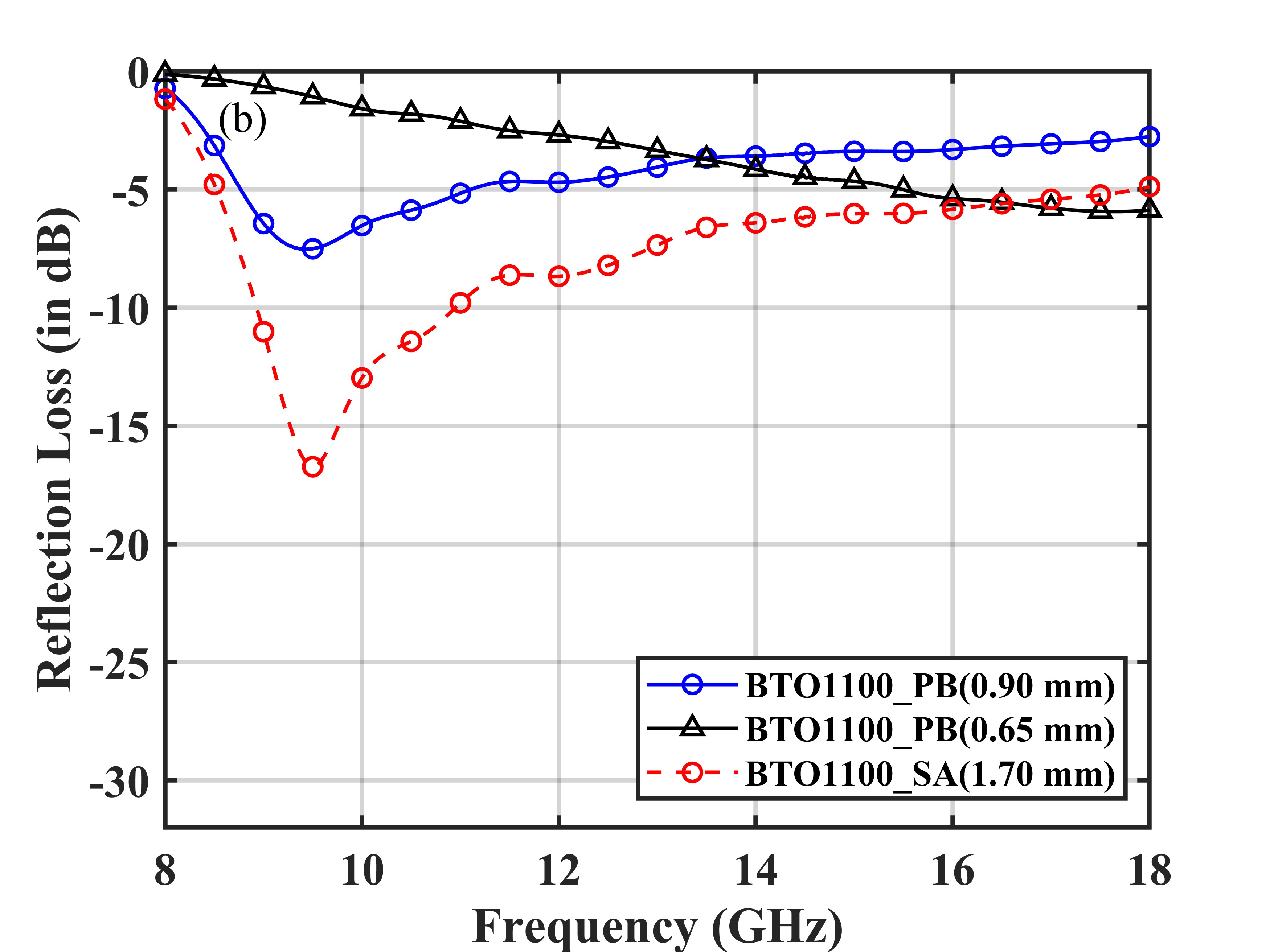}\label{fig:betterRLBTO1100_1} }}
  \captionsetup{labelsep=period}
  \caption{Comparison of the RL spectra of Pure block and Star auxetic samples made up of BTO1000 (Barium Titanate annealed till 1000$^{\circ}$C) and BTO1100 (Barium Titanate annealed till 1100$^{\circ}$C), where Star auxetic samples exhibited better RL capabilities. (a) BTO1000: 1.00 mm and 1.90 mm thick, (b) BTO1100: 0.90 mm and 1.70 mm thick. The plots in black colour represent the RL spectra of PB samples that have the same weight as the SA samples.}
\label{fig:betterRLBTO1000}
\end{figure}

From the previous section, it is evident that SA samples can provide different RL spectrum compared to PB samples. To extend the study carried out in the previous subsection, in the current subsection, the focus is to attain thickness pairs with the objective of identifying SA samples that can provide the weight advantage, enhanced minimum RL and bandwidth compared to the PB samples. 
There might be cases where PB might outperform in terms of RL with respect to SA; however, we are focusing only on cases where SA outperforms PB.
This section highlights the dual benefit of the SA architecture: enabling the utilisation of lighter designs in the absorber samples without compromising the RL capabilities and, in fact, enhancing the EM absorption performance. 
From the results discussed in this section, the possibility of simultaneously achieving increased efficacy in electromagnetic absorption and considerable weight savings using SA samples proves the appropriateness of such samples in the context of aerospace applications.
To have a stronger understanding of the advantages of using SA samples, the RL plots of the PB samples that have a similar weight to the chosen SA samples are also included in the Figure \ref{fig:betterRLBTO700},  Figure \ref{fig:betterRLBTO1000} and shown in black. 

For the BTO700 sample, the SA designs show better RL performance than the PB equivalents. The SA samples achieve a much lower minimum RL and a broader absorption band. From Figure \ref{fig:betterRLBTO7001}, the SA sample shows an RL of -23.43 dB and a bandwidth of 3.24 GHz, compared to the PB sample's -12.37 dB and 3.00 GHz. SA sample is able to provide weight savings of approximately 30\%.
For the BTO900 sample, SA continue to show superior EM absorption. The PB sample achieves -7.06 dB minimum RL and 1.91 GHz bandwidth, while the SA sample shows -15.00 dB and 4.30 GHz bandwidth (refer Figure \ref{fig:betterRLBTO9001}), with 31.9\% weight saving. 

The advantage of SA samples is more evident with BTO1000 (refer Figure \ref{fig:betterRLBTO1000_1}). The PB sample has shown EM absorption with minimum RL values close to -10 dB and a bandwidth of 3.68 GHz. The corresponding SA sample showed enhanced minimum RL ($<$ -20 dB) and bandwidth ($>$ 4.00 GHz), with weight savings of approximately 27\%.
With the BTO1100 sample, the SA again show better EM absorption capabilities (refer Figure \ref{fig:betterRLBTO1100_1}). The PB sample has a minimum RL of -7 dB with a bandwidth close to 2.30 GHz in the X-band. The corresponding SA sample maintains better RL performance with minimum RL$<$-15 dB and enhanced bandwidths exceeding 6.25 GHz in the 8-18 GHz range, along with substantial weight savings.

\textcolor{black}{These results demonstrate the two-fold advantage of SA architecture: enhanced EM absorption and weight savings across all BTO material configurations considered in the study. The potential of auxetic geometries is clearly shown for achieving lightweight, high-performance absorption.
Tables \ref{table:PB_SA_annealing_comparison_X_band} and \ref{table:PB_SA_annealing_comparison_Ku_band} consolidate all PB and SA pair results, showing the superior performance of SA samples in minimum RL, bandwidth, and weight savings. The results are separated into X and Ku bands for a better understanding of the EM absorption capabilities. All the RL comparison plots of the cases shown in the table are also plotted but not shown here.}

\begin{table}[!ht]
\centering
\caption{Comparison of thickness, minimum RL, bandwidth, and weight savings of PB and SA samples, where the RL spectra of SA samples are better than the PB samples in the X-band frequency range.}
\begin{tabular}{ccccccccc}
\hline
\multirow{2}{*}{\shortstack{Annealing \\Temperature}} & \multicolumn{2}{c}{Thickness (mm)} & \multicolumn{2}{c}{RL (dB)} & \multicolumn{2}{c}{BW (GHz)} & \multirow{2}{*}{Weight Saving (\%)} \\ \cline{2-7}
 & PB & SA & PB & SA & PB & SA & \\ \hline
\multirow{2}{*}{700$^{\circ}$C} 
    & 2.95 & 5.35 & -00.78 & -01.42 & 0.00 & 0.00 & 30.4 \\
    & 3.25 & 6.00 & -01.26 & -02.15 & 0.00 & 0.00 & 29.1 \\ \hline
\multirow{2}{*}{900$^{\circ}$C} 
    & 2.65 & 4.70 & -02.69 & -03.23 & 0.00 & 0.00 & 31.9 \\
    & 3.30 & 5.90 & -08.49 & -20.82 & 1.14 & 1.68 & 31.3 \\ \hline
\multirow{2}{*}{1000$^{\circ}$C} 
    & 1.00 & 1.90 & -11.57 & -29.53 & 3.13 & 3.60 & 26.9 \\
    & 1.10 & 2.15 & -11.43 & -22.85 & 2.87 & 4.00 & 27.3 \\ \hline
\multirow{2}{*}{1100$^{\circ}$C} 
    & 0.90 & 1.70 & -07.54 & -16.73 & 2.31 & 3.50 & 27.4 \\
    & 0.95 & 1.85 & -09.95 & -29.72 & 2.30 & 3.79 & 25.1 \\ \hline
\end{tabular}
\label{table:PB_SA_annealing_comparison_X_band}
\end{table}

\begin{table}[!ht]
\centering
\caption{Comparison of thickness, minimum RL, bandwidth, and weight savings of PB and SA samples, where the RL spectra of SA samples are better than the PB samples in the Ku-band frequency range.}
\begin{tabular}{ccccccccc}
\hline
\multirow{2}{*}{\shortstack{Annealing \\Temperature}} & \multicolumn{2}{c}{Thickness (mm)} & \multicolumn{2}{c}{RL (dB)} & \multicolumn{2}{c}{BW (GHz)} & \multirow{2}{*}{Weight Saving (\%)} \\ \cline{2-7}
 & PB & SA & PB & SA & PB & SA & \\ \hline
\multirow{2}{*}{700$^{\circ}$C} 
    & 2.95 & 5.35 & -12.40 & -24.13 & 3.06 & 3.24 & 30.4 \\
    & 3.25 & 6.00 & -12.48 & -23.80 & 2.25 & 3.33 & 29.1 \\ \hline
\multirow{2}{*}{900$^{\circ}$C} 
    & 2.65 & 4.70 & -07.04 & -15.00 & 1.91 & 4.30 & 31.9 \\
    & 3.30 & 5.90 & -03.41 & -07.56 & 0.00 & 0.44 & 31.3 \\ \hline
\multirow{2}{*}{1000$^{\circ}$C} 
    & 1.00 & 1.90 & -05.41 & -08.70 & 0.55 & 2.51 & 26.9 \\
    & 1.10 & 2.15 & -03.51 & -05.12 & 0.00 & 0.14 & 27.3 \\ \hline
\multirow{2}{*}{1100$^{\circ}$C} 
    & 0.90 & 1.70 & -04.66 & -08.67 & 0.00 & 4.00 & 27.4 \\
    & 0.95 & 1.85 & -04.14 & -06.77 & 0.00 & 2.46 & 25.1 \\ \hline
\end{tabular}
\label{table:PB_SA_annealing_comparison_Ku_band}
\end{table}

The EM performance of the proposed SA configuration with the ceramic-based metastructures reported in the recent literature for high-temperature applications is also shown in Table \ref{table:materials_comparison}. 
It can be seen from the Table \ref{table:materials_comparison} that the proposed SA configurations are less thick and show good RL performance. It is also to be noted that higher thickness configurations may be prove to catastrophic failure in the engine nozzle application due to higher thermal expansion and brittle nature of ceramics.


\begin{table}[!ht]
\centering
\caption{\textcolor{black}{Comparison of electromagnetic wave absorption performance of metamaterials made up of ceramic-based materials for high temperature applications.}}
\color{black}
\resizebox{1.00\textwidth}{!}{\begin{tabular}{ccccccc}
\hline
\multirow{2}{*}{Material} & \multirow{2}{*}{Minimum RL (dB)} & \multirow{2}{*}{Thickness (mm)} & \multirow{2}{*}{Specific RL (dB/mm)} & \multicolumn{2}{c}{BW (GHz)} & \\ \cline{5-6}
 & & & & X-band & Ku-band & \\ \hline
SiC \cite{wang2024material} & -46.00 & 18.0 & -02.56 & 4.0 & 4.0 & \\
Al$_2$O$_3$/SiC \cite{mei20193d} & -63.65 & 03.5 & -18.19 & 4.0 & - & \\
SicN \cite{pan2022high} & -49.00 & 12.2 & -04.02 & 4.0 & - & \\
MoS2/PyC-Al$_2$O$_3$ \cite{liu2023structural} & -18.00 & 06.5 & -02.77 & 4.0 & 6.0 & \\
Al$_2$O$_3$/SiC$_{\text{nw}}$/SiOC \cite{mei20213d} & -44.35 & 02.5 & -17.74 & 3.6 & - & \\
SiOC \cite{zhou2022digital} & -36.33 & 02.9 & -12.53 & 4.0 & - & \\
PyC/Al$_2$O$_3$ \cite{zhou20243d} & -38.00 & 03.0 & -12.67 & 4.0 & 6.0 & \\
SiC \cite{zhou2023stereolithographically} & -45.00 & 34.0 & -01.32 & 4.0 & 6.0 & \\
BTO1000\_SA & -29.53 & 01.9 & -15.54 & 3.6 & 2.5 & \\ 
BTO1100\_SA & -16.73 & 01.7 & -09.84 & 3.5 & 4.0 & \\ \hline
\end{tabular}}
\label{table:materials_comparison}
\end{table}

\subsection{Radar cross-section of aircraft embedded with PB and SA with enhanced RL and bandwidth in the nozzle region}
\label{sec:betterRLperformance}
\begin{figure}[!ht]
    \centering
      \subfloat{{\includegraphics[width=0.5\textwidth]{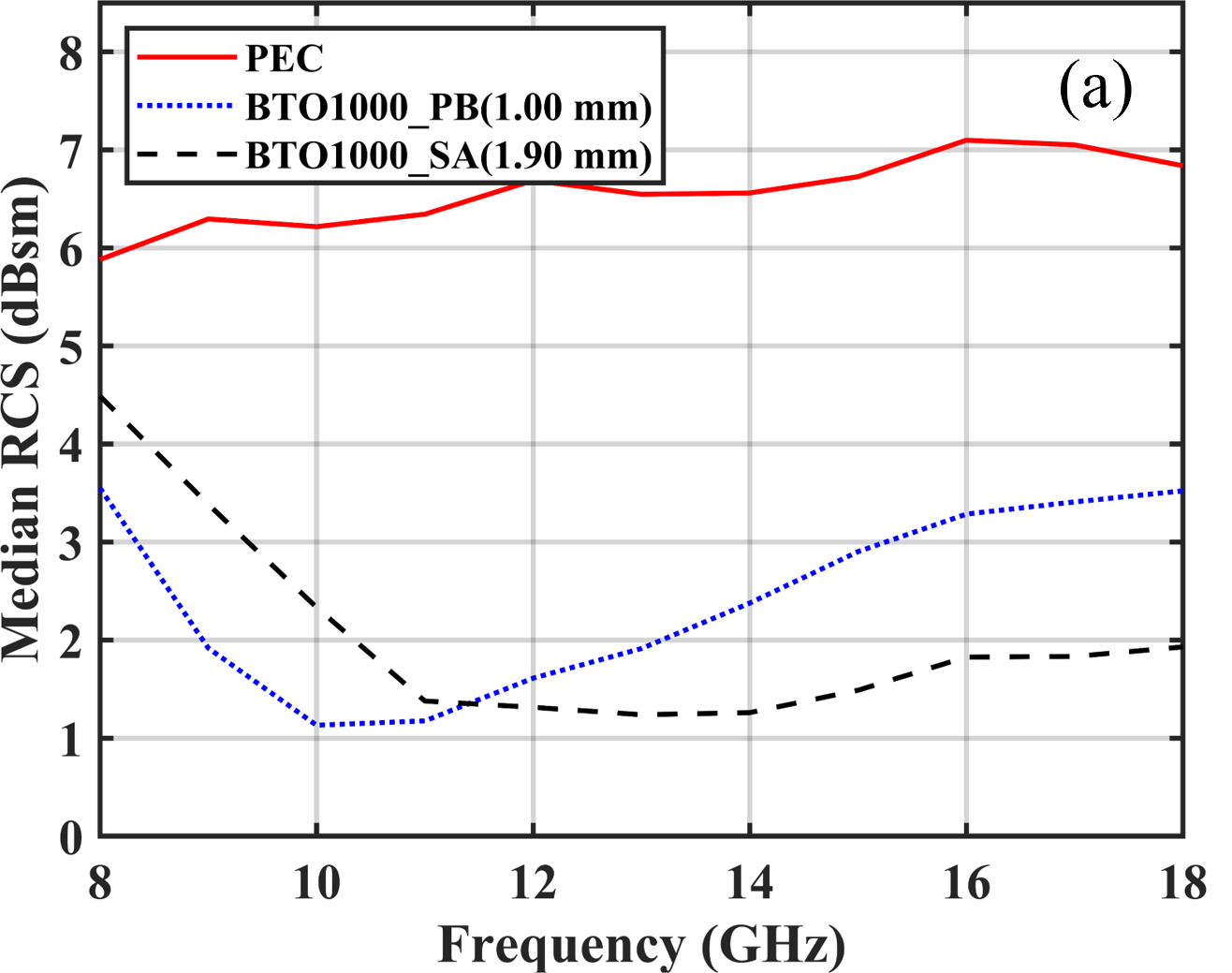}\label{MedianRCS1mm1.90mmHH} }}
  \subfloat{{\includegraphics[width=0.5\linewidth]{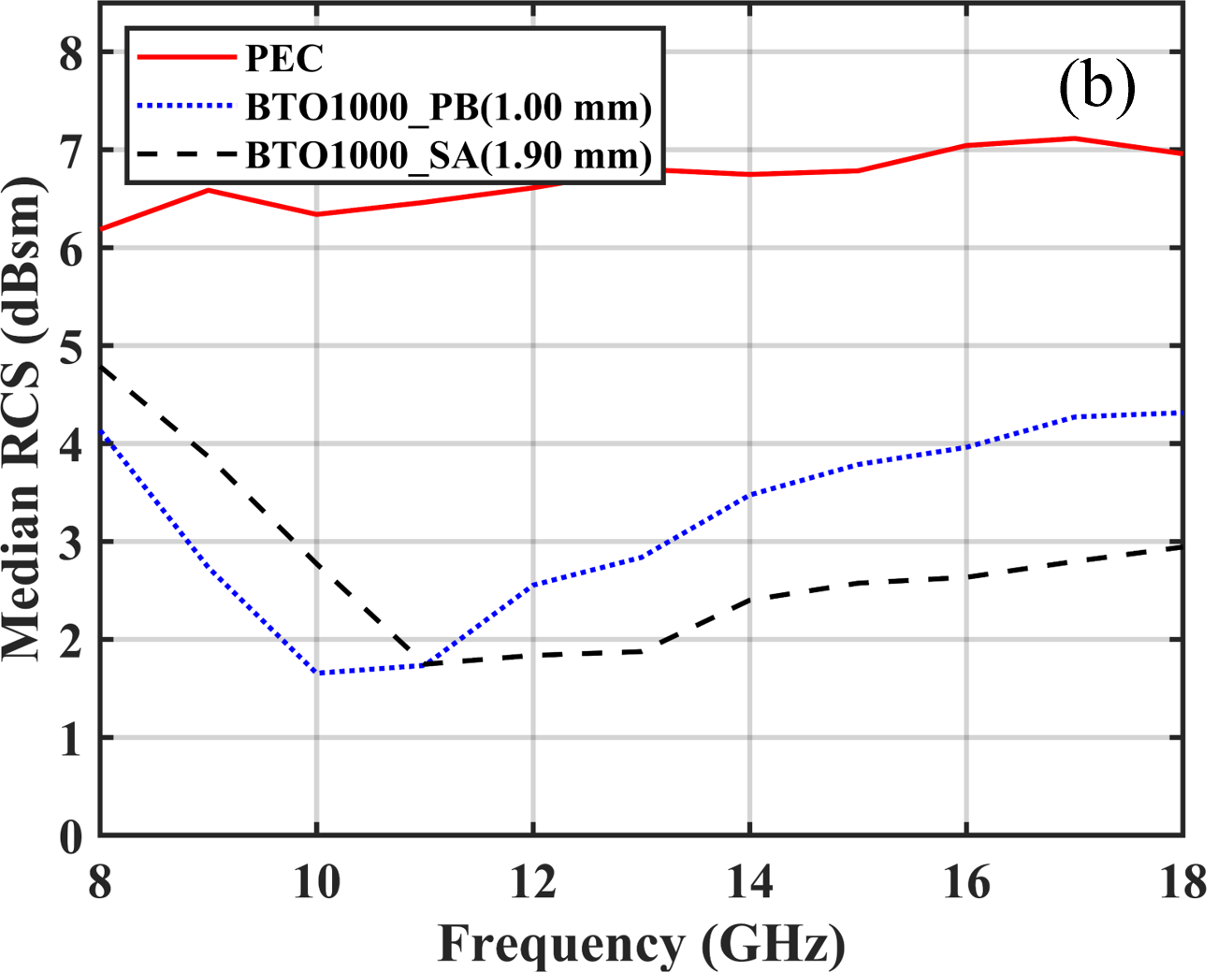}\label{MedianRCS1mm1.90mmVV}}}\\
        \subfloat{{\includegraphics[width=0.5\textwidth]{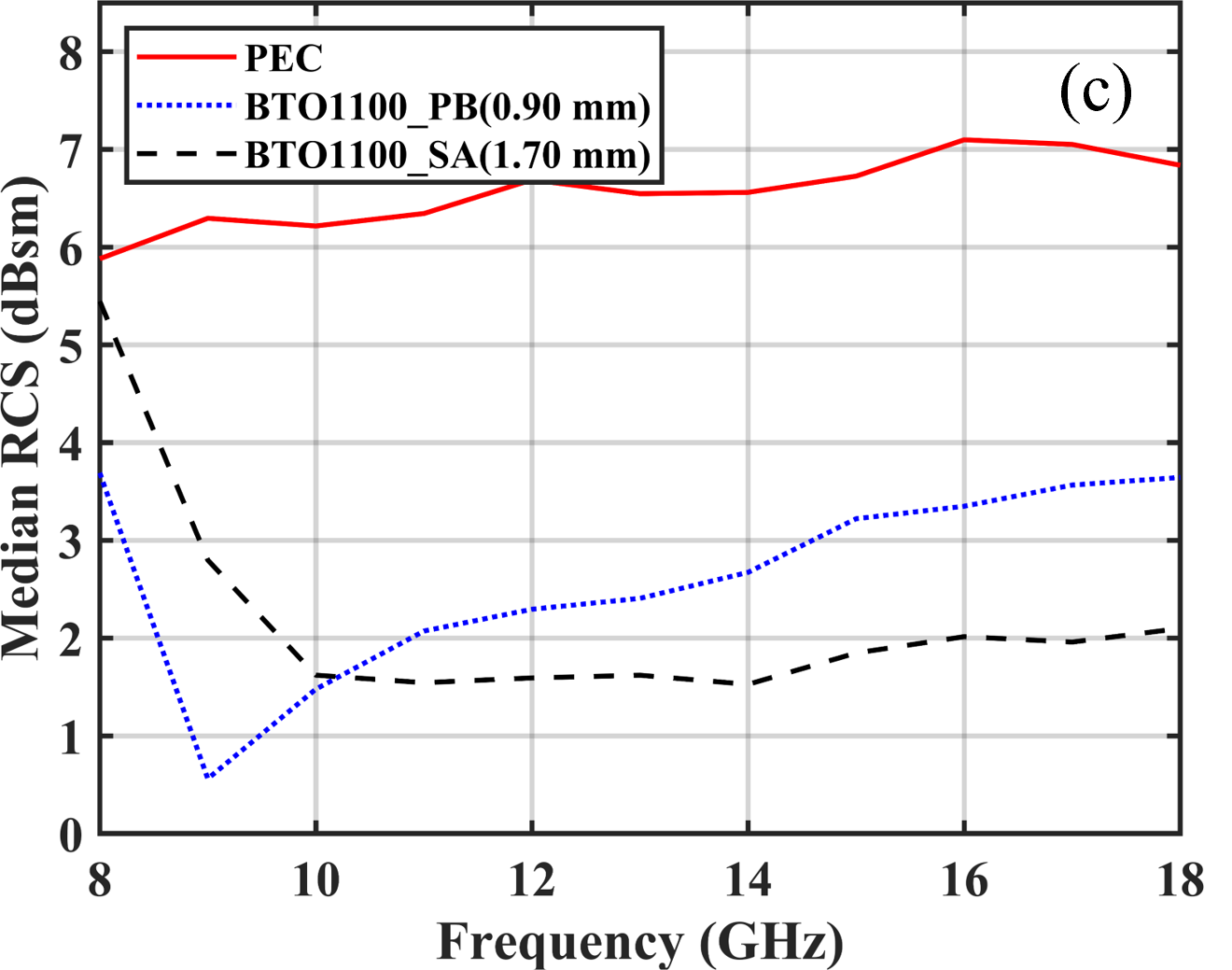}\label{MedianRCS0.9mm1.70mmHH} }}
  \subfloat{{\includegraphics[width=0.5\linewidth]{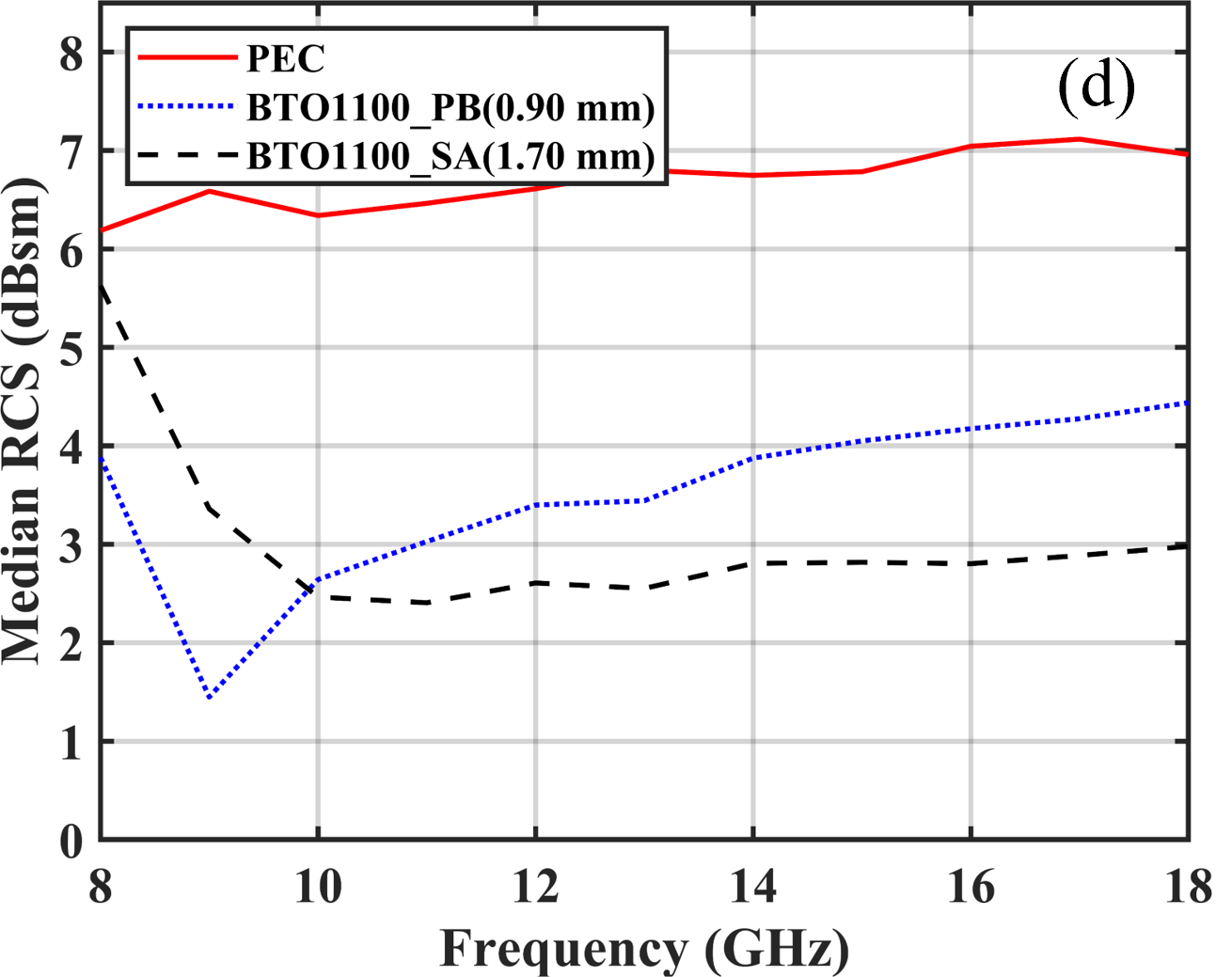}\label{MedianRCS0.9mm1.70mmVV}}}
  \captionsetup{labelsep=period}
  \caption{Rear aspect median RCS of aircraft variation in the frequency range of 8-18 GHz for horizontal (HH) and vertical (VV) polarisations. Results of the PEC nozzle comparison with (a) PB 1.00 mm and SA 1.90 mm nozzles for HH polarisation, (b) PB 1.00 mm and SA 1.90 mm nozzles for VV polarisation, (c) PB 0.90 mm and SA 1.70 mm nozzles for HH polarisation and (d) PB 0.90 mm and SA 1.70 mm nozzles for VV polarisation.}
\label{fig:MedianRCS1mm1.90mm0.90mm1.70mm}
\end{figure}

In this section, the rear aspect RCS performance of aircraft (nozzle with and without BTO) is evaluated. RCS performance of a few selected configurations from the considered SA-PB pairs in the Section \ref{sec:betterRLPBSA}, where SA designs have shown enhanced RL in comparison with the PB, are evaluated. The weight reduction of the BTO1000-based SA design shown in this section is approximately 27\% compared to the corresponding PB sample considered. The median RCS variation across X and Ku bands for HH and VV polarisations is shown in the Figure \ref{fig:MedianRCS1mm1.90mm0.90mm1.70mm}. 
Similar to the RL spectrum of PB 1.00 mm and SA 1.90 mm (refer to Figure \ref{fig:betterRLBTO1000_1}), it can be observed that the median RCS is also better for the SA. 
Between 8 and 18 GHz, the approximate difference in the median RCS between SA 1.90 mm and PEC is minimum at 8 GHz, which is 2.38 dB and maximum at 10 GHz, which is 5.10 dB.
The median RCS difference between SA 1.90 mm and PB 1.00 mm is a maximum of 1.56 dB at 18 GHz.
Similar to the RL spectrum of PB 0.90 mm and SA 1.70 mm (refer to Figure \ref{fig:betterRLBTO1100_1}), it can be observed that the median RCS is also better for the SA for the major portion in the 8-18 GHz frequency range. 
The approximate difference in the median RCS between SA 1.70 mm and PEC is minimum at 8 GHz, which is 0.50 dB and maximum at 12 GHz, which is 5.19 dB.
The median RCS difference between SA 1.70 mm and PB 0.90 mm is a maximum of 1.69 dB at 8 GHz.
Along with the cases discussed in this section, RCS plot variation for the other parametric studies are also studied but shown here for the sake of brevity.
\label{sec:RCSevaluation}
\section{Conclusions}
\label{sec:conclusions}
The in-house tools used in the study are initially validated by comparing the RL spectra with the experimental results from literature and Floquet port analysis carried out in the commercial software.
It is observed that the proposed combined tool is computationally efficient.
Four different parametric studies are carried out by varying the thicknesses of both SA and PB samples from 1.00 mm to 10.00 mm in increments of 0.05 mm, resulting in 181 unique configurations for each sample type. Pairwise comparisons are performed between all possible SA–PB combinations, yielding a total of 32,761 (181 $\times$ 181) cases. The observations from the studies are as follows:
\begin{itemize}
    \item When the same thickness samples at different BTO material configurations are compared, the SA configuration provides a decisive advantage in terms of weight saving up to 61.7\%. In many cases, SA offers superior RL and RCS reduction for specific BTO configurations and thickness.
    \item When compared at the same weight, certain SA configurations have shown better RL and RCS capabilities in comparison with the PB.
    \item When sample RL performance samples are compared, the SA configuration achieves comparable radar absorption performance to the PB with substantially less weight, demonstrating weight savings ranging from approximately 26\% to over 80\% depending on different BTO samples.
    \item Improved median RCS reduction throughout or major portions of the X and Ku bands, along with weight saving close to 30\%, is observed when SA samples, which can simultaneously provide enhanced RL, broader bandwidth, and significant weight savings over the PB samples, are considered.
\end{itemize}
Extensive comparative studies show that selective SA configurations can be employed on the aircraft nozzle and are able to provide superior RCS and RL in the X and Ku bands, compared to PB and PEC nozzles. There are cases in the selective SA configurations which have a thickness higher than the corresponding PB configuration with lower overall weight. Hence, the trade-off between the lower weight and higher thickness needs to be taken into account based on the application of interest.
Further RCS performance improvement with the same BTO material can be done by changing the shapes of nozzles to serpentine, S-shaped, and non-circular shapes, as a future scope of the work. 
\textcolor{black}{The fabrication of these complex-shaped auxetic ceramic components will also be explored using advanced manufacturing techniques such as Ultrafast Shaping and Sintering (USS), which allows for rapid, energy-efficient production of geometrically complex ceramics without compromising their functional properties.}
The proposed auxetic metamaterials, in combination with optimised nozzle profiles, provide the ability to open new ground in the creation of next-generation multifunctional stealth systems.
\section*{Declaration of competing interest}
The authors declare that they have no known competing financial
interests or personal relationships that could have appeared to influence the work reported in this paper.

\section*{Author Contribution Statement}
\textbf{A. Phanendra Kumar}: Conceptualisation, Methodology, Formal analysis, Investigation, Validation, Writing- Original draft preparation. 
\textbf{Preeti Kumari}: Formal analysis, Validation.
\textbf{Dineshkumar Harursampath}: Supervision, Writing-Review \& editing, Resources, Project administration. 
\textbf{Vijay Kumar Sutrakar}: Conceptualisation, Methodology, Supervision, Investigation, Writing-Review \& editing, Resources, Project administration.


\section*{Data availability}
Data cannot be shared due to the ongoing research activities.

\bibliographystyle{elsarticle-num}
\bibliography{bibliography}
\newpage
\appendix

\end{document}